\PassOptionsToPackage{most}{tcolorbox}
\documentclass[sigconf]{acmart}

\usepackage{tabularray}
\usepackage{tabularx}
\usepackage{hyperref} 
\usepackage{tikz}
\usepackage[table]{xcolor} % in the preamble

\usepackage{booktabs}
\usepackage{xcolor} % Ensure this is in your preamble
\usepackage{url}
\usepackage{paralist} % Used to have in paragraph lists
\usepackage{tcolorbox} % Package to generate colored box with text
\usepackage{graphicx} 
\usepackage{multirow}
\usepackage{makecell} % For better cell customization
\usepackage{xspace}
\usepackage{balance}
\usepackage{caption} % Used for multiple captions in a table
\usepackage{listings, xcolor}
\usepackage{relsize} % For relative font size adjustments

\lstdefinelanguage{Solidity}{
	keywords=[1]{anonymous, assembly, assert, balance, break, call, callcode, case, catch, class, constant, continue, constructor, contract, debugger, default, delegatecall, delete, do, else, emit, event, experimental, export, external, false, finally, for, function, gas, if, implements, import, in, indexed, instanceof, interface, internal, is, length, library, log0, log1, log2, log3, log4, memory, modifier, new, payable, pragma, private, protected, public, pure, push, require, return, returns, revert, selfdestruct, send, solidity, storage, struct, suicide, super, switch, then, this, throw, transfer, true, try, typeof, using, value, view, while, with, addmod, ecrecover, keccak256, mulmod, ripemd160, sha256, sha3}, % generic keywords including crypto operations
	keywordstyle=[1]\color{blue}\bfseries,
	keywords=[2]{address, bool, byte, bytes, bytes1, bytes2, bytes3, bytes4, bytes5, bytes6, bytes7, bytes8, bytes9, bytes10, bytes11, bytes12, bytes13, bytes14, bytes15, bytes16, bytes17, bytes18, bytes19, bytes20, bytes21, bytes22, bytes23, bytes24, bytes25, bytes26, bytes27, bytes28, bytes29, bytes30, bytes31, bytes32, enum, int, int8, int16, int24, int32, int40, int48, int56, int64, int72, int80, int88, int96, int104, int112, int120, int128, int136, int144, int152, int160, int168, int176, int184, int192, int200, int208, int216, int224, int232, int240, int248, int256, mapping, string, uint, uint8, uint16, uint24, uint32, uint40, uint48, uint56, uint64, uint72, uint80, uint88, uint96, uint104, uint112, uint120, uint128, uint136, uint144, uint152, uint160, uint168, uint176, uint184, uint192, uint200, uint208, uint216, uint224, uint232, uint240, uint248, uint256, var, void, ether, finney, szabo, wei, days, hours, minutes, seconds, weeks, years},	% types; money and time units
	keywordstyle=[2]\color{teal}\bfseries,
	keywords=[3]{block, blockhash, coinbase, difficulty, gaslimit, number, timestamp, msg, data, gas, sender, sig, value, now, tx, gasprice, origin},	% environment variables
	keywordstyle=[3]\color{violet}\bfseries,
	identifierstyle=\color{black},
	sensitive=true,
	comment=[l]{//},
	morecomment=[s]{/*}{*/},
	commentstyle=\color{gray}\ttfamily,
	stringstyle=\color{red}\ttfamily,
	morestring=[b]',
	morestring=[b]"
}

\lstdefinestyle{diff}{
    escapechar=\%
}

\lstdefinelanguage{prompt}{
  basicstyle=\footnotesize\ttfamily,
  literate={\$1}{{\textcolor{blue}{\bfseries\$1}}}2
           {\$2}{{\textcolor{blue}{\bfseries\$2}}}2
           {\$3}{{\textcolor{blue}{\bfseries\$2}}}2
           {poco}{{\textcolor{blue}{\bfseries{poco}}}}2
}

\usepackage{algorithm}
\usepackage{algpseudocode}

\usepackage{pifont}% http://ctan.org/pkg/pifont
\usepackage{float} % listing not split over multiple pages
\usepackage{listings}
\usepackage{minted}
\usepackage{xcolor}

\usepackage[export]{adjustbox}
\colorlet{numb}{magenta!60!black}

\definecolor{verylightgray}{rgb}{.97,.97,.97}
\definecolor{lightgreen}{rgb}{0.839,0.961,0.839}

\usepackage{minted}
\definecolor{darkgreen}{rgb}{0.0, 0.5, 0.0}
\newcommand{\cmark}{\textcolor{darkgreen}{\ding{51}}}%
\newcommand{\xmark}{\textcolor{red}{\ding{55}}}%
\newtcolorbox{answerbox}[1]{
    colback=black!5,              % slightly stronger background
    colframe=black!50,            % stronger contrast
    coltitle=black,
    title={#1},
    fonttitle=\bfseries,
    attach title to upper,
    after title={:\ },
    borderline west={1.6pt}{0pt}{black!65}, % stronger accent bar
    boxrule=0.4pt,                % subtle full border
    arc=2mm,
    width=0.96\linewidth,
    center
}

\newtcolorbox{promptbox}{
  enhanced,
  colback=black!2,          % lighter than findings
  colframe=black!25,        % softer border
  boxrule=0.3pt,            % thinner than findings
  arc=2pt,
  left=6pt,
  right=6pt,
  top=4pt,
  bottom=4pt,
  fontupper=\ttfamily\small,
}

\definecolor{codegray}{rgb}{0.95,0.95,0.95}

\definecolor{gold}{RGB}{184,134,11}

\usepackage{fontawesome5}
\newcommand{\success}{{\color{gold}\faTrophy}}

\renewcommand{\texttt}[1]{%
    {\ttfamily\hyphenchar\font=45\relax #1}%
}
\newcommand{\tool}{\textsc{PoCo}\xspace}
\newcommand{\maxcost}{$\$3$\xspace}

\newcommand{\datasetsize}{$23$\xspace}
\newcommand{\dataset}{\textsc{Proof-of-Patch}\xspace}
\newcommand{\rqone}{To what extent can \tool generate well-formed PoC exploits for smart contracts? \xspace}
\newcommand{\rqtwo}{To what extent can \tool generate logically correct PoC exploits for smart contracts? \xspace}
\newcommand{\rqthree}{What impact do different levels of detail in vulnerability annotations have on the results?\xspace}

\newcommand{\openai}{OpenAI o3\xspace}
\newcommand{\claude}{Claude Sonnet 4.5\xspace}
\newcommand{\glm}{GLM 4.6\xspace}

\begin{document}

%\title{A Comprehensive Study of Automated Exploit Mitigation for Smart Contracts}
%\title{Does Automatic Fixing of Smart Contract Vulnerabilities Actually Mitigate Exploits?}
\title{PoCo: Agentic Proof-of-Concept Exploit Generation for Smart Contracts}

\author{Vivi Andersson}
\authornote{Both authors contributed equally to this research.}
\email{vivia@kth.se}
\orcid{0009-0000-6519-625X}
\affiliation{%
  \institution{KTH Royal Institute of Technology}
  \city{Stockholm}
  \country{Sweden}
}
\author{Sofia Bobadilla}
\email{sofbob@kth.se}
\orcid{0000-0003-3116-3278}
\authornotemark[1]
\affiliation{%
  \institution{KTH Royal Institute of Technology}
  \city{Stockholm}
  \country{Sweden}
}

\author{Harald Hobbelhagen}
\email{hhob@kth.se}
\orcid{0009-0001-1963-7525}
\affiliation{%
  \institution{KTH Royal Institute of Technology}
  \city{Stockholm}
  \country{Sweden}
}

\author{Martin Monperrus}
\orcid{0000-0003-3505-3383}
\email{monperrus@kth.se}
\affiliation{%
  \institution{KTH Royal Institute of Technology}
  \city{Stockholm}
  \country{Sweden}
}

\begin{abstract}
Smart contracts operate in a highly adversarial environment, where vulnerabilities can lead to substantial financial losses. 
Thus, smart contracts are subject to security audits.
In auditing, proof-of-concept (PoC) exploits play a critical role by demonstrating to the stakeholders that the reported vulnerabilities are genuine, reproducible, and actionable. 
However, manually creating PoCs is time-consuming, error-prone, and often constrained by tight audit schedules.
%what is PoCo
We introduce \tool, an agentic framework that automatically generates executable PoC exploits from natural-language vulnerability descriptions written by auditors.
%how it operates
\tool autonomously generates PoC exploits in an agentic manner by interacting with a set of code-execution tools in a Reason–Act–Observe loop. It produces fully executable exploits compatible with the Foundry testing framework, ready for integration into audit reports and other security tools.

%results
We evaluate \tool on a dataset of 23 real-world vulnerability reports. 
\tool consistently outperforms the Zero-shot and Workflow baselines, generating well-formed and logically correct PoCs. 
Our results demonstrate that agentic frameworks can significantly reduce the effort required for high-quality PoCs in smart contract audits.
Our contribution provides actionable knowledge for the smart contract security community.
\end{abstract}

\maketitle

% Note that keywords are not normally used for peer review papers.
% \begin{IEEEkeywords}
%IEEE, IEEEtran, journal, \LaTeX, paper, template.
%\end{IEEEkeywords}

\newcommand{\TODO}[1]{\textcolor{red}{#1}\GenericWarning{}{LaTeX Warning: TODO: #1}}
\newcommand{\todo}{\TODO}

% \fi

%\def\thefootnote{1}\footnotetext{Equal contribution
%}\def\thefootnote{\arabic{footnote}}
\section{Introduction}
\label{sec:introduction}
% Problem statement
Smart contracts operate in an extremely adversarial environment. 
As of October 2025, on-chain exploits have resulted in approximately \$15 billion in losses~\cite{defillama_hacks}.
With new exploits emerging daily, there is an urgent need to prevent vulnerable contracts from being deployed. If they are deployed, they will be exploited.

The smart contract lifecycle typically comprises four stages \cite{lifecycle2025}: (1) development and testing, (2) security auditing, (3) deployment, and (4) monitoring. 
Security auditing involves third-party experts assessing the smart contract code through careful manual and tool-supported analysis, and reporting vulnerabilities to the smart contract development team.
Audits are essential for ensuring a project’s reliability and security, yet limited budgets and tight timelines often constrain them.

When writing an audit, it is a best practice to report a clear description of the vulnerability, an impact assessment, and a proof-of-concept (PoC) exploit~\cite{immunefiPocGuidelines2024}.
% define PoC
A PoC demonstrates that the reported issue can produce harmful behavior, such as asset loss or protocol malfunction.

PoCs are critical: for auditors, they provide verifiable evidence of a vulnerability; for developers, they serve as actionable test cases to reproduce and fix the issue; and for stakeholders, they offer clear evidence for risk assessment and prioritization.
Despite their critical role, manually creating PoCs is time-consuming and error-prone, often constrained by tight audit schedules \cite{Feist2023EvaluatingBlockchainSecurityMaturity}. 
Previous work on automating this process generates blockchain transaction sequences~\cite{teether, maian, foray2024, flashsyn2024, wei2025veriexploit} rather than executable tests, focuses solely on vulnerabilities with direct monetary extraction~\cite{gervais2025ai}, or generates exploits based on a fixed pipeline~\cite{xiao2025prompt2pwn} that cannot adapt to real-world smart contract software projects.
The problem we address in this paper is the automated generation of PoCs from high-level textual descriptions.

This paper introduces \tool, an agentic framework that generates executable Foundry test code from natural-language descriptions, for real-world Solidity smart contract protocols.
\tool takes as input the target smart contract and an auditor-written vulnerability annotation, and produces ready-to-execute PoC exploits.
By automating the PoC generation, vulnerabilities can be documented more thoroughly and remediated faster, reducing the risk of exploitable contracts reaching production.

\tool's design is a state-of-the-art agentic loop, combining full autonomy and domain-specific tools related to smart contract development. We use \tool with three frontier LLMs: \claude, \openai, and \glm.

We evaluate \tool on 23 real-world vulnerabilities from smart contract projects. 
We first evaluate the capabilities of \tool at generating well-formed PoC exploits that compile (RQ1): \tool generates executable PoCs in 50 runs.
Next, we evaluate whether the PoCs are correct, with an original methodology based on the ground-truth mitigation patch (RQ2): \tool generates 32 correct PoC exploits for our real-world vulnerabilities.
For perspective, we compare \tool against two baselines: Zero-shot Prompting and Workflow Prompting.
To evaluate the sensitivity to quality of vulnerability annotations, we assess the performance when varying their level of descriptive detail (RQ3).
Across all measures, \tool achieves the highest success rates in generating both well-formed and logically correct PoCs, by a large margin.
To our knowledge, we are the first to report on real-world PoC exploit generation at this scale and level of difficulty, with only real-world vulnerabilities.

To summarize, we make the following contributions:
\begin{itemize}
    \item \tool, a blueprint of real-world proof-of-concept exploit generation for smart contracts based on agentic AI.
    \item A novel evaluation methodology that uses developer-provided mitigation patches as independent correctness oracles for automated PoC validation.
    \item An experimental campaign on 23 real-world smart contract vulnerabilities and 3 models, demonstrating that \tool generates three times more logically correct PoCs than workflow baselines. The results are available at \href{https://github.com/ASSERT-KTH/PoCo-public/}{github.com/ASSERT-KTH/PoCo-public/}.
    \item \dataset, a high-quality dataset based on security audits that links vulnerabilities, PoCs, and their mitigation patches. The dataset is fully reproducible and open-source at \href{https://github.com/ASSERT-KTH/Proof-of-Patch}{https://github.com/ASSERT-KTH/proof-of-patch}.
\end{itemize}

The remainder of this paper is organized as follows. Section~\ref{sec:background} provides the necessary background. Section~\ref{sec:tool} presents \tool. Section~\ref{sec:experimental_methodology} describes our evaluation methodology and Section~\ref{sec:experimental_results} the corresponding results. Section~\ref{sec:discussion} discusses limitations and threats to validity. Section~\ref{sec:related_work} covers related work, and Section~\ref{sec:conclusion} concludes.

\section{Background}\label{sec:background}
Our work lies at the intersection of smart contract security and AI for code. This section covers key concepts for understanding our contribution and its novelty.

\subsection{Blockchain Concepts}

\paragraph{Smart Contract \& Solidity.}
Smart Contracts are programs deployed on blockchain networks to enforce predefined rules without the need for intermediaries. 
Contracts are written in high-level smart contract languages, such as Solidity, and then compiled into bytecode for deployment and execution. 
A smart contract deployed on-chain is immutable; this means its code cannot be modified or altered.
Solidity is Ethereum’s primary smart contract programming language. 
From a security perspective, the misuse of constructs such as fallback functions and low-level calls can lead to security vulnerabilities.

\paragraph{Transactions.}
All operations on the blockchain are initiated via transactions. 
A transaction is typically sent from a user, signed by their private key, to perform actions such as transferring cryptocurrency or invoking functions on existing contracts.
During execution, a called contract may interact with other contracts.

\paragraph{Lifecycle of Smart Contract Development }\label{sec:sc-lifecycle}
The lifecycle begins with the development phase, where protocol developers implement and test the smart contract logic. 
Development is followed by a security review, also known as a smart contract audit (see Section~\ref{concept:audit}). Developers can then fix the security problems identified in the audit.
Once done, the contract is deployed on a production blockchain. 
Finally, there is a post-deployment monitoring phase to ensure ongoing security and functionality through real-time surveillance~\cite{your_exploit_is_mine}, anomaly detection, and automated reporting \cite{lifecycle2025}.

% on the importance of audits EARLY in the life cycle: "A late lifecycle penetration testing paradigm uncovers problems too late, at a point when both time and budget severely constrain the options for remedy. I" (software penetration testing, arkins, classic https://ieeexplore.ieee.org/stamp/stamp.jsp?arnumber=1392709)

\paragraph{Smart Contract Audits}\label{concept:audit}
The purpose of an audit is to systematically identify vulnerabilities, logical errors, and design flaws, thereby preventing exploits and ensuring the protocol functions as intended. 
This process is a cornerstone for ensuring the security and reliability of blockchain applications. 
Projects engage external security firms or independent experts to perform these audits.

\begin{figure*}
    \centering
    \includegraphics[width=\linewidth]{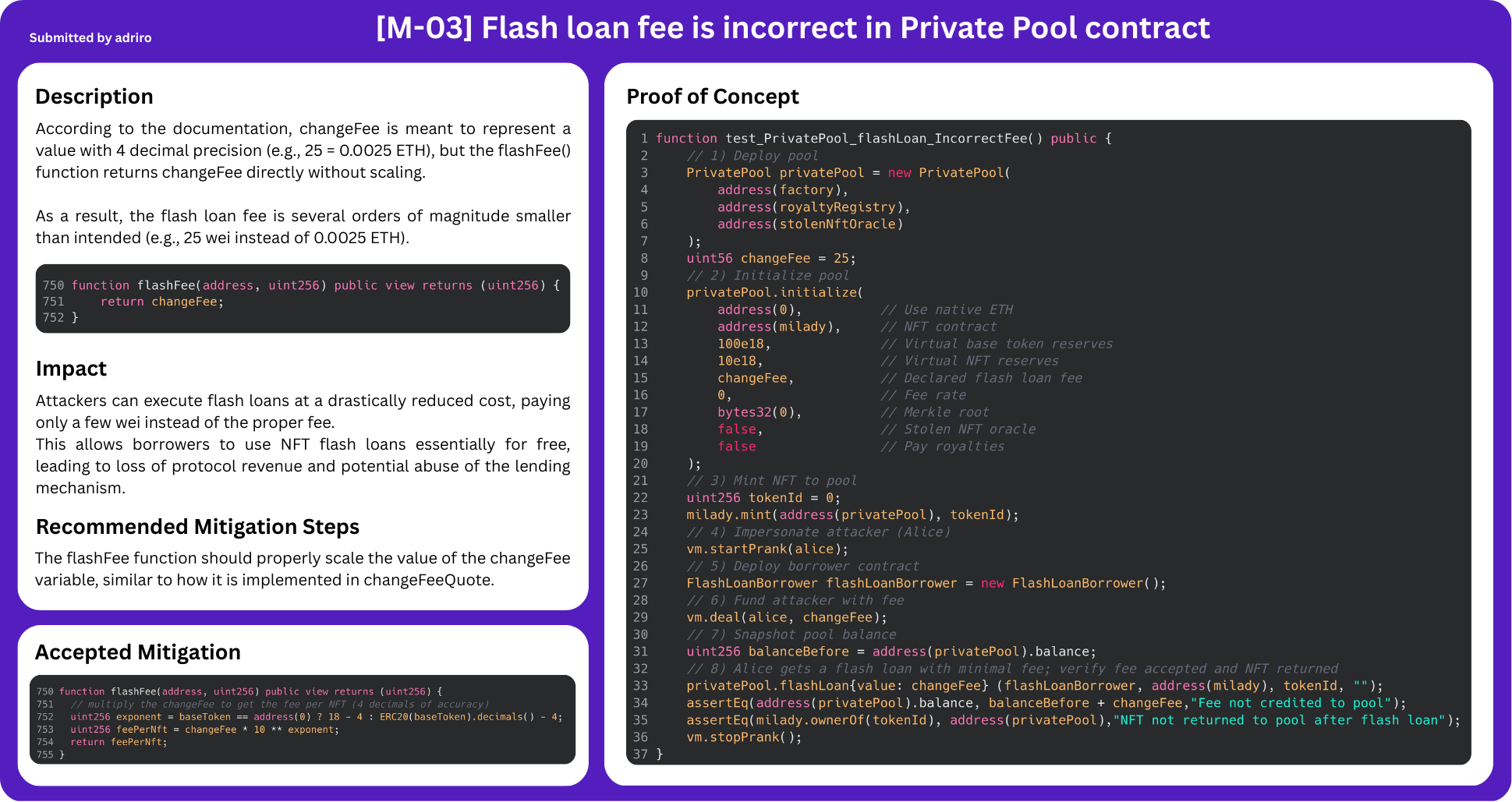}
    \caption{Vulnerability example, inspired by the audit competition code4rena 2023-04-caviar.}
    \label{fig:finding_example}
\end{figure*}

\paragraph{Audit Competition Platforms}
Audit competition platforms such as Code4rena\footnote{\href{https://code4rena.com/audits}{https://code4rena.com/}} and Immunefi\footnote{\href{https://immunefi.com/blog/all/immunefi-audits/}{https://immunefi.com/}} employ a crowdsourced security model where projects open their codebase to all security researchers for time-bound competitions. 
Participants compete to identify vulnerabilities, submitting detailed reports for monetary rewards scaled by the severity and quality of their findings. 
A valid \textit{finding} must demonstrate a specific, exploitable issue in the code that could lead to security compromises. 
Audit competitions are public, hence can be used to collect rich security datasets that provide foundational material for data-driven security research, including AI-based approaches. Figure \ref{fig:finding_example} showcases a real-world vulnerability finding from an audit competition platform.

\paragraph{PoC Definition}\label{sec:definitions}
% this is framework specific
A proof-of-concept exploit is an executable demonstration that shows that a claimed vulnerability can be triggered under controlled conditions.
A PoC demonstrates that a vulnerability, in theory, can be exploited in practice. 
A white-hat PoC validates vulnerability impact without live exploitation on-chain \cite{immunefiPocRequired}.
This is crucial for blockchain environments where live testing could irrevocably compromise immutable contracts and tangible assets.
PoC exploits play a crucial role for both auditors and development teams. For auditors, they offer indisputable, executable proof that the reported vulnerability is genuine and reproducible. For the receiving (developer) team, they serve as actionable artifacts that can be used to reproduce, validate, and ultimately mitigate the underlying issue.
For reference, see  Proof of Concept on Figure~\ref{fig:finding_example}. %\todo{a few sentences about what we see in the figure}.
The protocol documentation specifies a flash loan fee of 0.0025 ETH; however, the implementation fails to enforce this rate. 
The PoC demonstrates how a user (represented as Alice in the code) can successfully execute a flash loan while paying an incorrect fee of only 25 wei, a negligible amount compared to 0.0025 ETH (1 ETH = $10^{18}$ wei). This discrepancy is validated by the assertions on lines 34-35 of the PoC.

\subsection{Agentic AI}
\textit{Agentic AI} refers to systems capable of autonomous goal pursuit through iterative cycles of planning, acting, observing, and revising~\cite{Ng2025welcoming}. 
Recent approaches achieve this autonomy by \emph{scaffolding} large language models (LLMs) with components for task decomposition, tool use, and memory~\cite{weng2023agent}. 
The autonomy of LLM-based agents varies from non-agentic single-turn prompting to systems that can formulate subgoals, invoke external tools, and iteratively refine their behavior toward a defined objective.

% AI for code
In the context of programming, \textit{agentic AI for code} refers to systems that can write, test, and iteratively improve software through various feedback cycles~\cite{roychoudhury2025agentic}. Such systems move beyond static code generation by autonomously identifying errors, revising solutions, and validating outcomes.

\subsection{Challenges of PoC Construction}

Generating a PoC is non-trivial: it requires bridging the gap between a suspected vulnerability and an executable exploit scenario~\cite{bobadilla2025automated}. 
Identifying a vulnerable code location is insufficient; one must demonstrate that 1) it can be reached under realistic execution conditions and 2) it leads to a concrete security violation ~\cite{vulnotexploit}.

We categorize these challenges into those inherent to PoC construction in the smart contract domain and those that arise when the task is delegated to an LLM agent.

\subsubsection{Domain Challenges}

\paragraph{(1) Semantic Implicitness.}
Constructing a PoC requires showing that a flaw can be triggered in a way that violates a security property.
This entails satisfying path conditions, state invariants, and execution constraints that are implicit in the codebase and rarely made explicit in vulnerability reports. % atoo much agent issue?

\paragraph{(2) Operational Complexity.}
PoCs rarely consist of a single function call~\cite{efcf2023}.
They require staged initialization, state manipulation, and coordinated multi-transaction interactions across contracts.
Hence, in a working PoC, contracts must be deployed and initialized correctly, and transactions built in the proper order.
Any deviation prevents the exploit from materializing.

For example, the vulnerability report in Figure~\ref{fig:finding_example} first requires setting up the scenario for a valid flash loan invocation (operational complexity) within lines 2--32. The exploit is then triggered at line 33 through the flash loan execution, and finally validated in lines 34--35, where the security violation is asserted (semantic implicitness).

\subsubsection{AI Challenges}
When PoC construction is delegated to a large language model, the domain challenges above manifest as three concrete failure modes.

\paragraph{(3) Well-formedness.}
The model must generate code that compiles, deploys, and executes successfully.
Compilation failures, incorrect dependencies, malformed transactions, and improper initialization are common.
Even when the vulnerability is correctly understood, small syntactic or environmental errors prevent the exploit from materializing.

\paragraph{(4) Correctness.}
The generated PoC must demonstrate the intended security violation.
Due to semantic implicitness, a model may produce plausible-looking and executable code that reaches the vulnerable function yet fails to satisfy the preconditions required to trigger the flaw.
Logical correctness depends on whether the generated execution trace aligns with the vulnerability specification, not merely on successful execution.

\paragraph{(5) Input Sensitivity.}
Natural-language vulnerability reports vary widely in style and procedural granularity.
Some provide detailed reproduction guidance, while others describe only the high-level issue and its impact.
As procedural detail decreases, the burden on the model to infer missing semantic and operational constraints increases, directly affecting both well-formedness and correctness.

\noindent To the best of our knowledge, no state-of-the-art automated tools address this problem for real-world smart contracts  (see Section~\ref{sec:related_work}).
This is where \tool makes a contribution.

\section{\tool}\label{sec:tool}
% Explain what the tool is, what problem it solves, and how it fits in the paper's overall goal.

\tool implements an autonomous single LLM agent system to generate PoC exploits. In this paper, we define a single LLM agent as an LLM operating tools in a goal-directed loop.
%REF https://simonwillison.net/2025/Sep/18/agents/

\subsection{Overview}
The goal of \tool is to synthesize executable PoC exploits for vulnerabilities that smart contract auditors have just identified.
% sentence on why this is important (what problem it solves).
Executable PoCs that demonstrate a vulnerability are crucial for comprehending its severity and impact on the targeted protocol \cite{immunefiPocGuidelines2024}. 

\begin{figure*}
     \centering
     \includegraphics[width=\linewidth]{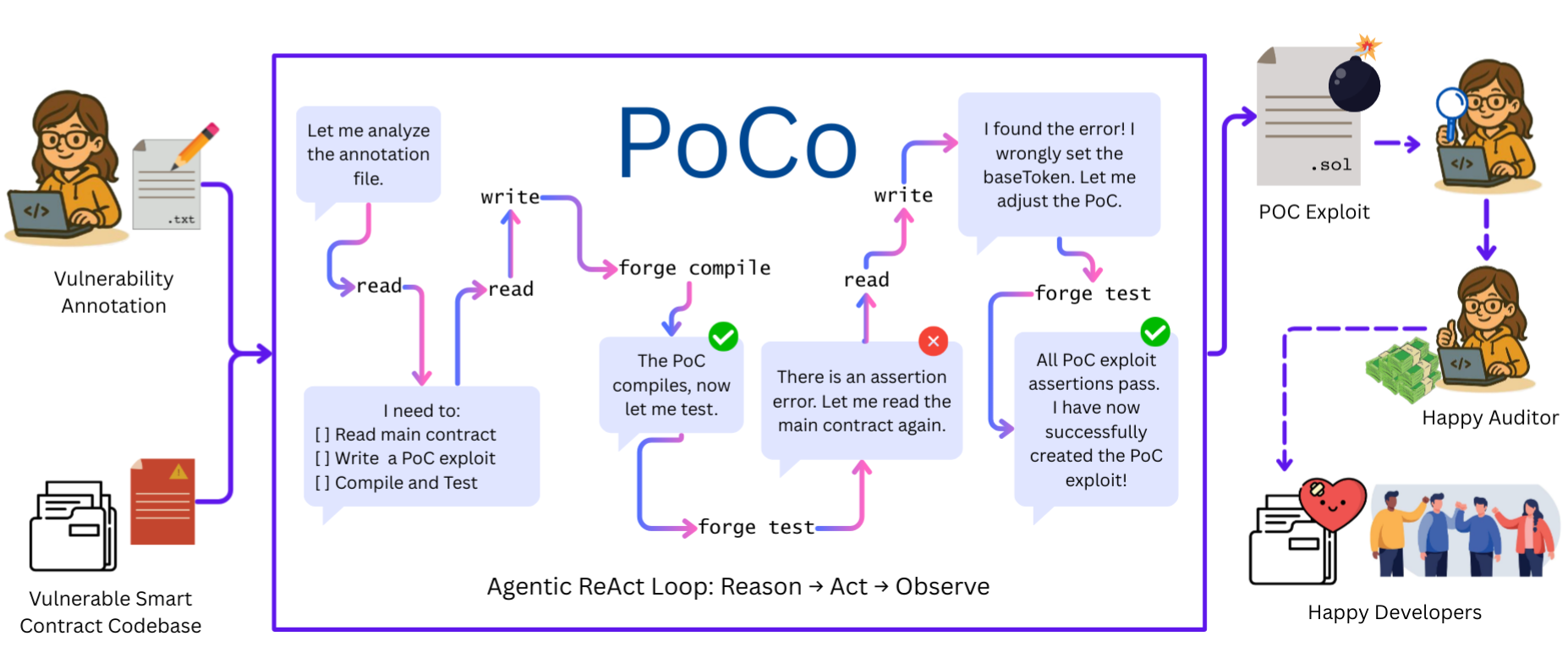}
     \caption{Overview of \tool's workflow. Starting with a smart contract project under audit and a vulnerability annotation written by an auditor, \tool autonomously crafts an executable PoC exploit. The auditor reviews and validates the PoC before submitting the vulnerability report, potentially receiving a monetary reward. The development team receives the PoC along with reproduction steps, enabling them to reproduce the issue and implement a patch efficiently.}
     \label{fig:poco-overview}
\end{figure*}

Figure~\ref{fig:poco-overview} gives a high-level overview of the workflow. Given a vulnerable contract and a description of the vulnerability in natural language written by the auditor, \tool synthesizes a PoC (Sec.~\ref{sec:definitions}) with assertions that expose the vulnerability. 
After receiving the PoC exploit, the auditor conducts a manual verification to confirm its validity. Then, the auditor can submit the vulnerability report with an increased likelihood of receiving a higher financial reward, thanks to the presence of the PoC.

%(Sec. \ref{sec:definitions}). %(Sec.~\ref{sec:sc-lifecycle}).
% Clarify the role of tool and preconditions
Note that \tool's scope is only to generate PoC exploits, not to discover vulnerabilities.
% \tool operates in a pre-deployment scenario with full local access to 1) the project sources 2) the project test environments and 3) a local blockchain environment. 
The considered development workflow involves the auditor performing manual analysis and writing a vulnerability annotation in natural language, in accordance with field auditing practices \cite{lifecycle2025}.

% summarise what the figure depicts (data flow, main components)

\subsection{Agentic Architecture}\label{sec:agentic-arch}
% high level principles, one llm facts, on llm tools, defined, which executes code . Key design principles: (1) (2) 3
\tool adopts an agentic architecture that couples a single LLM with a set of external tools for compilation, execution, and exploration.
The model possesses full autonomy in selecting, ordering, and configuring tool calls, including control over command flags.
According to the bitter lesson principle \cite{sutton2019bitter}, we minimize scaffolding; the surrounding scaffold primarily exists to expose the tool interfaces and manage input–output flow. This ensures that \tool does not overfit to a single model, and is flexible to tool changes (addition, modification).

In practice, \tool operates through a small number of core tools:

\subsubsection{Basic Tools}\label{sec:basic-tools}
The basic toolset provides \tool with the ability to explore and modify its working environment. 
These include file-system search (\verb|glob|, \verb|grep|), file reading (\verb|read|), and file editing or writing (\verb|edit|, \verb|write|). 
This minimal set supports inspection and modification of project files throughout the analysis process.

\subsubsection{Planning Tool}
For internal task organization, \tool makes use of a lightweight planning utility, exposed through a \verb|Todo| interface that supports reading and updating task entries. The utility is general-purpose and domain-agnostic. It helps the agent to track progress over a previously elaborated plan.

\subsubsection{Smart Contract Tools}\label{sec:domain-tools}
We define and provide access to two tools specific to smart contracts: \verb|smart-contract|
\verb|-compile| compiles Solidity contracts, and \verb|smart-contract-test| executes the generated PoC exploit in the vulnerable smart contract codebase.
To implement those tools, we select Foundry’s Forge framework\footnote{\href{https://getfoundry.sh/forge/overview}{https://getfoundry.sh/forge/overview}} for its maturity and its widespread adoption in modern Solidity-based smart contract projects\footnote{As of 2024, Foundry is the most popular Solidity framework \href{https://www.soliditylang.org/blog/2025/04/25/solidity-developer-survey-2024-results/}{https://www.soliditylang.org/blog/2025/04/25/solidity-developer-survey-2024-results/}}.
These tools enable the agent to validate the PoC exploits it generates.
The tool output also provides \tool with critical feedback, allowing \tool to self-correct previously produced PoCs. 

%The commands are exposed as MCP tools rather than Bash commands. This provides stronger guardrails against tool-use errors \cite{yang2024swe}. 
All commands are executed within a containerized environment to (1) mitigate potential security risks of running exploit code, and (2) to ensure reproducibility of exploit executions.
\verb|smart-contract|
\verb|-compile| and \verb|smart-contract-test| produce raw outputs which are parsed into a standardized result schema capturing the command’s execution status, output, and error streams.

\subsection{Agent Command}\label{sec:agent-command}
% general strategy
\tool follows a ReAct-style ~\cite{yao2022react} loop of reasoning, acting, and observing, enabling it to decide on subsequent actions through environment feedback autonomously. 

To achieve the goal of synthesizing a valid PoC exploit, we command \tool through our defined system prompt and task prompts.
The prompts were developed by iteratively monitoring agent behavior on a held-out development dataset not included in our evaluation data.

% SYSTEM
The system prompt describes the agent's role and expected behavior. The latter is a description of a blueprint approach to generate PoC exploits. It is grounded in advice from established smart contract PoC best practices~\cite {immunefiPocGuidelines2024}. Furthermore, it contains information about:

\begin{enumerate}
      \item \textbf{PoC Explainability.} Generate a PoC that is an \textit{executable demonstration} of the vulnerability and clearly document all attack steps.
      \item \textbf{Vulnerability analysis guidelines:} Parse the provided vulnerability description to understand the affected code, root cause, and potential impact.
      \item \textbf{Framework-specific guidelines:} Employ Foundry-specific features, including dedicated state setup and syntax (e.g., \verb|vm.prank|).
      \item \textbf{PoC Executability guidelines:} Verify that the PoC exploit \textit{compiles} and its tests \textit{pass}.
      \item \textbf{Iterative Refinement Hints:} Iterate on the code to resolve any compilation, test, or logical errors.
      \item \textbf{Exploit Soundness Criteria:} Produce a logically sound exploit that accurately reflects the analyzed vulnerability.
      \item \textbf{Exploit Quality Enforcement:} Keep the PoC minimal, avoiding modifications to existing source code and adding helpers only when strictly necessary.
\end{enumerate}
The full system prompt is found in Appendix~\ref{app:system-prompt}.

% \begin{figure}[t]
% \begin{minipage}{0.95\linewidth}
% \begin{lstlisting}[
%   caption={Agent task instruction template},
%   label=lst:task-prompt,
%   breaklines=true,
%   language=prompt,
%   numbers=none
% ]
% Create a vulnerability exposing PoC forge test for the vulnerable contract at $1 using the vulnerability description in $2. Use the Write tool to save your PoC code to $3. Write ONLY the test file, test ONLY the described vulnerability, and do NOT modify the original contract. Iterate on compilation, test, and logical errors using forge tools. You are done when the test compiles and successfully demonstrates the vulnerability through passing assertions.
% \end{lstlisting}
% \end{minipage}
% \end{figure}
\begin{figure*}[t]
\centering
\captionsetup{type=lstlisting} % treat next caption as a listing
\begin{tcolorbox}[
    colback=gray!5,
    colframe=black!75,
    width=\textwidth,
    arc=2mm,
    boxrule=0.5pt,
    left=6pt,
    right=6pt,
    top=6pt,
    bottom=6pt
]
{\small
\noindent\textbf{Task Prompt:}

\medskip
\noindent
Create a vulnerability exposing PoC forge test for the vulnerable contract at \texttt{\$1} using the vulnerability description in \texttt{\$2}. Use the Write tool to save your PoC code to \texttt{\$3}. Write ONLY the test file, test ONLY the described vulnerability, and do NOT modify the original contract. Iterate on compilation, test, and logical errors using forge tools. Your task is finished when the test compiles and successfully demonstrates the vulnerability through passing assertions.
}
\end{tcolorbox}
\caption{\tool task prompt for PoC generation.}
\label{lst:task-prompt}
\end{figure*}

% \begin{promptbox}
% Create a vulnerability exposing PoC forge test for the vulnerable contract at \texttt{\$1} using the vulnerability description in \texttt{\$2}. Use the Write tool to save your PoC code to \texttt{\$3}. Write ONLY the test file, test ONLY the described vulnerability, and do NOT modify the original contract. Iterate on compilation, test, and logical errors using forge tools. You are done when the test compiles and successfully demonstrates the vulnerability through passing assertions.
% \end{promptbox}

A task prompt is then appended to the system prompt. It is a minimal instruction about generating a PoC exploit, shown in Listing~\ref{lst:task-prompt}.
Then, we pass three file paths to the agent: (1) the path to the vulnerable contract, (2) the path to the auditor's vulnerability annotation, and (3) the target path for the PoC exploit file to be written. 

\subsection{Agent Behavior Monitoring}
To enable post-hoc analysis of agent performance, we implement comprehensive monitoring of \tool to capture two complementary aspects: (1) complete interaction trajectories and (2) execution metadata: tokens, cost, tool calls, messages.

\textbf{Interaction trajectories:}
\tool records the full sequence of messages happening during the agent interaction.
These trajectories capture prompts, assistant responses, user messages, tool invocations with their parameters and return values, error states, and generated text outputs. This complete trace captures the decision-making context necessary for understanding the agentic behavior.

\textbf{Execution metadata:}
\tool saves metadata related to resource consumption and model configuration. Resource consumption metrics include token counts, cumulative model costs, total session duration, and conversation round counts. Model metadata captures the specific model identifier and the temperature setting. Additionally, we record execution times for individual tool calls to identify performance bottlenecks.

\subsection{Guardrails}
A smart contract agent system with code-execution capabilities could potentially perform unintended or harmful actions, such as interacting with real on-chain contracts or accessing sensitive host data. 
To prevent this, we restrict \tool's action space in two main ways. 
First, the system interacts only with an allowlist of CLI commands, none of which permit sending transactions on-chain. 
Second, all code execution occurs within isolated Docker containers, preventing the agent from accessing sensitive host content such as internal password files or SSH keys. 
These guardrails ensure that \tool can safely generate proofs-of-concept exploits without causing harmful side effects on real-world mainnets. %NOTE we are running privileged containers though, is an issue?

% Dreadnode: There are plenty of valid reasons for restricting the action space, including reliability, predictability, and safety. However, these restrictions ultimately prevent the models from being able to respond to situations the restricted tooling doesn’t allow for. This means a human must intervene to upgrade the API with the necessary change to access new capabilities.

\subsection{Implementation}
Our agent scaffolds the Claude Code SDK (Anthropic, v2.0.10).
Domain-specific capabilities are exposed as MCP servers implemented with FastMCP (v.2.12.3); these servers run inside Docker (version 23.0.3) containers, which in turn run Foundry (version 1.3.1) to provide isolation and reproducibility. Model routing and selection are performed by the Claude Code router (v1.0.53), which forwards model calls to OpenRouter API endpoints. A CLI wrapper encapsulates the SDK calls and routing configuration.

% Infrastructure:

% Docker containerization (isolation benefits, reproducibility)
% Hosting environment (local vs. cloud)
% Latency considerations for tool execution
% Resource constraints (memory, compute)

\section{Experimental Methodology}
\label{sec:experimental_methodology}

% Guidelines for Empirical Studies in Software Engineering involving Large Language Models https://arxiv.org/pdf/2508.15503
% (1) to declare LLM usage and role; (2) to report model versions, configurations, and fine-tuning; (3) to document tool architectures; (4) to disclose prompts and interaction logs; (5) to use human validation; (6) to employ an open LLM as a baseline; (7) to use suitable baselines, benchmarks, and metrics; and (8) to openly articulate limitations and mitigations

\subsection{Dataset}\label{sec:dataset}

\subsubsection{Properties}
Our goal is to evaluate whether the proof-of-concept exploits generated by \tool are both executable and logically correct.
To automatically assess the first, the PoC exploit is added to a smart contract project codebase and compiled.
For the second, the exploit is executed against the corresponding security patch, see Section~\ref{subsec:rq2-methodology}.

Achieving both requires a dataset with specific properties. 
Each entry in the dataset must contain:
\begin{enumerate}
    \item Real-world audit cases describing the presence and nature of the vulnerability.
    \item The corresponding vulnerability patch, either as part of the audit report or as a pull request link.
    \item The corresponding smart contract project source code.
    \item An executable Foundry project configuration, including test cases.
\end{enumerate}

\subsubsection{Limitations of Existing Datasets}
There exist several well-known datasets for smart contract auditing and vulnerability analysis, including 
SC-Bench~\cite{sc-bench}, 
DAPPScan~\cite{DAppSCAN},  
ScaBench,\footnote{\href{https://github.com/scabench-org/scabench/}{github.com/scabench-org/scabench/. Accessed on October 4th.}} and 
Verite~\cite{verite}. 
However, none of them meet all the criteria aforementioned. 
SC-Bench focuses exclusively on ERC-standard contracts and does not capture the diversity of DeFi protocols.
DAPPScan provides real audit reports from different platforms; however, they lack a standard and structured format, making it infeasible for scalable data extraction.
% updated
ScaBench provides a curated benchmark for assessing automated end-to-end auditing tools. The dataset aggregates audit competitions to achieve broad vulnerability coverage. It does not, however, focus on mitigation, and therefore, patch availability is critically limited. % only 9 references to a PR from 555 samples
Finally, Verite~\cite{verite}, based on DeFiHackLabs\footnote{\href{https://github.com/SunWeb3Sec/DeFiHackLabs}{github.com/SunWeb3Sec/DeFiHackLabs. Accessed on 2025-10-04}}, contains post-deployment incidents, valuable for forensic studies but not for evaluating pre-deployment patch effectiveness.

\subsubsection{Data Collection}
To address this gap, we construct our own dataset, \dataset, by collecting real-world audit findings and linking them to executable source code and verified patches.
We source audit findings from Solodit\footnote{\href{https://solodit.cyfrin.io/}{https://solodit.cyfrin.io/. Accessed on 2025-10-04}} a centralized platform that aggregates public audit findings with metadata (incl. vulnerability, audit contest, impact, and associated code repository).
We query Solodit for all vulnerability categories defined in the OWASP Smart Contract Top 10\footnote{https://owasp.org/www-project-smart-contract-top-10/. Accessed on 2025-10-04}: 
\href{https://owasp.org/www-project-smart-contract-top-10/2025/en/src/SC01-access-control.html}{access control}, 
\href{https://owasp.org/www-project-smart-contract-top-10/2025/en/src/SC02-price-oracle-manipulation.html}{price oracle manipulation},
\href{https://owasp.org/www-project-smart-contract-top-10/2025/en/src/SC03-logic-errors.html}{logic errors},
\href{https://owasp.org/www-project-smart-contract-top-10/2025/en/src/SC04-lack-of-input-validation.html}{lack of input validation},
\href{https://owasp.org/www-project-smart-contract-top-10/2025/en/src/SC05-reentrancy-attacks.html}{reentrancy attacks},
\href{https://owasp.org/www-project-smart-contract-top-10/2025/en/src/SC06-unchecked-external-calls.html}{unchecked external calls},
\href{https://owasp.org/www-project-smart-contract-top-10/2025/en/src/SC07-flash-loan-attacks.html}{flash loan attacks},
\href{https://owasp.org/www-project-smart-contract-top-10/2025/en/src/SC08-integer-overflow-underflow.html}{integer overflow and underflow}, 
\href{https://owasp.org/www-project-smart-contract-top-10/2025/en/src/SC09-insecure-randomness.html}{insecure randomness}, and
\href{https://owasp.org/www-project-smart-contract-top-10/2025/en/src/SC10-denial-of-service.html}{denial of services}.
From every report, we collect: 
(i) project metadata such as the project name, report link, publication date, and authors; 
(ii) vulnerability details, including category, severity, and description; 
(iii) indicators of exploit evidence, such as the presence of a proof-of-concept, either as code, text, or an external reference; and 
(iv) indicators of mitigation, including links to commits or pull requests and textual or code-based descriptions of the proposed fix.  
This structured data enables us to perform tasks such as prioritization and manual analysis.

\subsubsection{Prioritization}
The raw collection yielded 3,814 audit findings, which were too numerous to inspect manually in full. 
Therefore, to assist manual evaluation, we assess the quality of the findings using the LLM-as-Judge methodology \cite{llm_as_judge}. 
A language model helps assess the quality of each report per the following dimensions: whether a patch is proposed, whether a PoC is included, whether a mitigation strategy is suggested, and an overall quality assessment. The quality assessment is categorized as \textit{bad}, \textit{fair}, \textit{good}, or \textit{excellent}.
We rank all samples in the raw dataset according to their scores. 

\subsubsection{Manual Analysis}
Using our automated prioritization, we then select the top 100 samples for manual validation. 
For each audit finding, we manually check whether
\begin{enumerate}
    \item The audit project is publicly available, executable, and properly configured for testing with Foundry.
    \item The audit judges have validated the finding.
    \item A valid patch exists for the finding.
\end{enumerate}

For each audit finding, we impose a one-hour time limit for installation and configuration. Samples that cannot be successfully set up within this time are discarded.
For findings deemed suitable for our project, we include the corresponding patch and store its description as the annotation. If the description contains a coded proof-of-concept, we store the PoC exploit for reference but exclude the code itself from the annotation.

\subsubsection{Proof of Patch}
The final dataset, \dataset, can be accessed at \hyperlink{https://github.com/ASSERT-KTH/Proof-of-Patch}{ASSERT-KTH/proof-of-patch}, and collects \datasetsize smart contracts projects.
It contains recent findings from 2022 to 2025-07.
Its characteristics are summarized in Table~\ref{tab:proof_of_patch_dataset}.
To the best of our knowledge, this is the first smart contract dataset that links valid vulnerability findings with the corresponding security patches implemented by Web3 developers.

\SetTblrInner{rowsep=1pt, colsep=2pt}

\begin{table*}
\centering
\begin{tblr}{
    colspec ={c l p{0.52\linewidth} p{0.06\linewidth} p{0.06\linewidth} p{0.05\linewidth} },
    column{2} = {rightsep = 4pt},
    column{3} = {rightsep = 6pt},
    column{4} = {leftsep  = 6pt},
    %row{even} = {gray!30!white},
    %has POC008,015,018,020, 033, 041,042,046,048,054,070,077,091
    row{4, 6, 7, 8, 10, 12, 13, 14, 15, 18, 21, 22, 23} = {gray!30!white},
    cells   = {font = \fontsize{9pt}{8pt}\selectfont},
}
\textbf{ID} & \textbf{Project} & \textbf{Description} &   \textbf{Audit  Ref.} &  \textbf{Patch \\ Ref.}& \textbf{Has \\ PoC} \\

    \href{https://solodit.cyfrin.io/issues/m-01-multicall-does-not-work-as-intended-code4rena-size-size-git}{001} &
    \href{https://github.com/code-423n4/2024-06-size}{2024-06-size} & 
    Logical error in multicall function allows users to bypass deposit limits.& 
    %Medium & 
    \href{https://github.com/code-423n4/2024-06-size-findings/issues/238}{M-01}  
    & \href{https://github.com/SizeCredit/size-solidity/pull/126}{\#PR126} & 
    No 
    \\
    \href{https://solodit.cyfrin.io/issues/h-04-vaultmintyieldfee-function-can-be-called-by-anyone-to-mint-vault-shares-to-any-recipient-address-code4rena-pooltogether-pooltogether-git}{003} & 
    \href{https://github.com/code-423n4/2023-07-pooltogether}{2023-07-pooltogether} &
    User can mint shares to any address and steal the yield fee of the protocol. &
    %High &
    \href{https://github.com/code-423n4/2023-07-pooltogether-findings/issues/396}{H-04} &
    \href{https://github.com/GenerationSoftware/pt-v5-vault/pull/7}{\#PR7} &
    No 
    \\
    \href{https://solodit.cyfrin.io/issues/m-05-investors-claiming-their-maxdeposit-by-using-the-liquiditypooldeposit-will-cause-other-users-to-be-unable-to-claim-their-maxdepositmaxmint-code4rena-centrifuge-centrifuge-git}{008} &
    \href{https://github.com/code-423n4/2023-09-centrifuge}{2023-09-centrifuge} &
    Rounding errors in share calculations allow investors to receive excess shares. &
    %Medium &
    \href{https://github.com/code-423n4/2023-09-centrifuge-findings/issues/118}{M-05}&
    \href{https://github.com/centrifuge/liquidity-pools/pull/166}{\#PR166} &
    Yes 
    \\
    \href{https://solodit.cyfrin.io/issues/m-08-loss-of-funds-for-traders-due-to-accounting-error-in-royalty-calculations-code4rena-caviar-caviar-private-pools-git}{009} &
    \href{https://github.com/code-423n4/2023-04-caviar}{2023-04-caviar} &
    Royalties are miscalculated when recipient address is zero, leading to trapped funds. &
    %Medium &
    \href{https://github.com/code-423n4/2023-04-caviar-findings/issues/596}{M-08} &
    \href{https://github.com/outdoteth/caviar-private-pools/pull/11/files}{\#PR11}&
    No 
    \\
    \href{https://solodit.cyfrin.io/issues/m-02-unintended-or-malicious-use-of-prize-winners-hooks-code4rena-pooltogether-pooltogether-git}{015} &
    \href{https://github.com/code-423n4/2023-07-pooltogether}{2023-07-pooltogether} & 
    The prize-winners hook mechanism can be exploited to interfere with the intended prize distribution process. &
    %High &
    \href{https://github.com/code-423n4/2023-07-pooltogether-findings/issues/465}{M-02} &
    \href{https://github.com/GenerationSoftware/pt-v5-vault/pull/21}{\#PR21}&
    Yes 
    \\
    \href{https://solodit.cyfrin.io/issues/m-15-pool-tokens-can-be-stolen-via-privatepoolflashloan-function-from-previous-owner-code4rena-caviar-caviar-git}{018} &
    \href{https://github.com/code-423n4/2023-04-caviar}{2023-04-caviar} &
    Former owner can set token approvals that enable them to reclaim assets after ownership transfer. &
    %High &
    \href{https://github.com/code-423n4/2023-04-caviar-findings/issues/230}{M-15}&
    \href{https://github.com/outdoteth/caviar-private-pools/pull/2}{\#PR2}&
    Yes 
    \\
    \href{https://solodit.cyfrin.io/issues/m-3-share-price-inflation-by-first-lp-er-enabling-dos-attacks-on-subsequent-buyshares-with-up-to-1001x-the-attacking-cost-sherlock-dodo-gsp-git}{020} & 
    \href{https://github.com/sherlock-audit/2023-12-dodo-gsp}{2023-12-dodo-gsp} &
    A first liquidity provider can inflate the share price during pool initialization, enabling a DoS. &
    %Medium &
    \href{https://github.com/sherlock-audit/2023-12-dodo-gsp-judging/issues/55}{M-03}&
    \href{https://github.com/DODOEX/dodo-gassaving-pool/pull/14/files}{\#PR14}&
    Yes 
    \\
    \href{https://solodit.cyfrin.io/issues/m-06-denial-of-service-contract-owner-could-block-users-from-withdrawing-their-strike-code4rena-putty-putty-contest-git}{032} &
    \href{https://github.com/code-423n4/2022-06-putty}{2022-06-putty} &
    User cannot withdraw their strike amount and their asset will be stuck in the contract. &
    %Medium &
    \href{https://github.com/code-423n4/2022-06-putty-findings/issues/296\#issuecomment-1185411399}{M-06}&
    \href{https://github.com/outdoteth/putty-v2/pull/4/files}{\#PR4} &
    No 
    \\
    \href{https://solodit.cyfrin.io/issues/m-03-flash-loan-fee-is-incorrect-in-private-pool-contract-code4rena-caviar-caviar-private-pools-git}{033} &
    \href{https://github.com/code-423n4/2023-04-caviar}{2023-04-caviar} &
    The PrivatePool contract miscalculates flash loan fees causing incorrect fee totals. &
    %Medium &
    \href{https://github.com/code-423n4/2023-04-caviar-findings/issues/864}{M-03}&
    \href{https://github.com/outdoteth/caviar-private-pools/pull/6}{\#PR6}&
    Yes 
    \\
    \href{https://solodit.cyfrin.io/issues/h-2-m-1-sherlock-axis-finance-git}{039} &
    \href{https://github.com/sherlock-audit/2024-03-axis-finance}{2024-03-axis-finance} &
    Refund handling errors can lock seller funds when the token reverts on zero transfers.&
    %High &
    \href{https://github.com/sherlock-audit/2024-03-axis-finance-judging/issues/21}{M-01}&
    \href{https://github.com/Axis-Fi/axis-core/pull/142/files}{\#PR142}&
    No 
    \\
    \href{https://solodit.cyfrin.io/issues/h-1-malicious-user-can-overtake-a-prefunded-auction-and-steal-the-deposited-funds-sherlock-axis-finance-git}{041} &
    \href{https://github.com/sherlock-audit/2024-03-axis-finance}{2024-03-axis-finance}&
    User can hijack a prefunded auction and gain control over its deposited funds. &
    %High &
    \href{https://github.com/sherlock-audit/2024-03-axis-finance-judging/issues/12}{H-01}&
    \href{https://github.com/Axis-Fi/moonraker/pull/132}{\#PR132}&
    Yes 
    \\
    \href{https://solodit.cyfrin.io/issues/m-2-utilization-rate-multiplier-will-not-shift-if-oracle-is-consulted-frequently-sherlock-cap-git}{042} &
    \href{https://github.com/sherlock-audit/2025-07-cap}{2025-07-cap} &
    User can exploit a rounding error to repeatedly miscompute utilization, causing inaccurate interest rate adjustments. &
    %Medium &
    \href{https://github.com/sherlock-audit/2025-07-cap-judging/issues/148}{M-02}&
    \href{https://github.com/cap-labs-dev/cap-contracts/pull/187}{\#PR187}&
    Yes 
    \\
    %%%
    \href{https://solodit.cyfrin.io/issues/m-03-zero-token-transfer-can-cause-a-potential-dos-in-cvxstaker-code4rena-xeth-xeth-versus-contest-git}{046} &
    \href{https://github.com/code-423n4/2023-05-xeth}{2023-05-xeth} &
    Zero token transfer can cause a potential denial of service when giving rewards &
    %Medium &
    \href{https://github.com/code-423n4/2023-05-xeth-findings/issues/30}{M-03}&
    \href{https://github.com/code-423n4/2023-05-xeth/commit/1f714868f193cdeb472ec097110901a997d87ec4}{1f71a}&
    Yes 
    \\
    %%%
    \href{https://solodit.cyfrin.io/issues/h-01-royalty-receiver-can-drain-a-private-pool-code4rena-caviar-caviar-private-pools-git}{048} &
    \href{https://github.com/code-423n4/2023-04-caviar}{2023-04-caviar} &
    Malicious royalty recipient can extract value from the pool without proper payment. &
    %High &
    \href{https://github.com/code-423n4/2023-04-caviar-findings/issues/593\#issuecomment-1520075272}{H-01}&
    \href{https://github.com/outdoteth/caviar-private-pools/pull/12}{\#PR12}&
    Yes 
    \\
    \href{https://solodit.cyfrin.io/issues/m-2-lender-is-able-to-steal-borrowers-collateral-by-calling-rollloan-with-unfavourable-terms-on-behalf-of-the-borrower-sherlock-cooler-update-git}{049} &
    \href{https://github.com/sherlock-audit/2023-08-cooler}{2023-08-cooler} &
    Lender can update loan terms without borrower approval, enabling them to impose unfair conditions. &
    %Medium &
    \href{https://github.com/sherlock-audit/2023-08-cooler-judging/issues/26}{M-02}&
    \href{https://github.com/ohmzeus/Cooler/pull/54/files\#diff-f461174637e644b69004d9f7ad97d531a760909f465ad610acea335531a49767}{\#PR54}&
    No 
    \\
    %%%%
    \href{https://solodit.cyfrin.io/issues/m-04-you-can-deposit-really-small-amount-for-other-users-to-dos-them-code4rena-centrifuge-centrifuge-git}{051} &
    \href{https://github.com/code-423n4/2023-09-centrifuge}{2023-09-centrifuge} &
    Missed access control allows users to deposit on behalf of others and potentially caused a denial of service attack. &
    %Medium &
    \href{https://github.com/code-423n4/2023-09-centrifuge-findings/issues/143}{M-04}&
    \href{https://github.com/centrifuge/liquidity-pools/pull/136}{\#PR136}&
    No 
    \\
    %%%%%
    \href{https://solodit.cyfrin.io/issues/h-01-no-revert-on-transfer-erc20-tokens-can-be-drained-code4rena-cally-cally-contest-git}{054} &
    \href{https://github.com/code-423n4/2022-05-cally}{2022-05-cally} &
    Unchecked token transfer return values let attackers create empty vaults, causing buyers to pay Ether but receive no tokens. &
    %High &
    \href{https://github.com/code-423n4/2022-05-cally-findings/issues/89}{H-01}&
    \href{https://github.com/outdoteth/cally/pull/4}{\#PR4}&
    Yes 
    \\
    \href{https://solodit.cyfrin.io/issues/m-05-fillorder-and-exercise-may-lock-ether-sent-to-the-contract-forever-code4rena-putty-putty-git}{058} &
    \href{https://github.com/code-423n4/2022-06-putty}{2022-06-putty} &
    Users can accidentally send Ether to code paths that don’t use it, causing the funds to be locked &
    %Medium &
    \href{https://github.com/code-423n4/2022-06-putty-findings/issues/226}{M-05}&
    \href{https://github.com/outdoteth/putty-v2/pull/5}{\#PR5}&
    No 
    \\
    \href{https://solodit.cyfrin.io/issues/h-02-protocol-mints-less-rseth-on-deposit-than-intended-code4rena-kelp-dao-kelp-dao-git}{066} &
    \href{https://github.com/code-423n4/2023-11-kelp}{2023-11-kelp} &
    Users receive less rsETH than expected due to a miscalculation in the minting logic. &
    %High &
    \href{https://github.com/code-423n4/2023-11-kelp-findings/issues/62}{H-02} &
    \href{https://github.com/code-423n4/2023-11-kelp-findings/issues/62\#issuecomment-1850480282}{Other}&
    No \\
    %%%%%%
    \href{https://solodit.cyfrin.io/issues/m-01-contract-phinft1155-cant-be-paused-code4rena-phi-phi-git}{070} &
    \href{https://github.com/code-423n4/2024-08-ph}{2024-08-ph} &
    Users are able to transfer NFT tokens even when the contract is paused. &
    %Medium &
    \href{https://github.com/code-423n4/2024-08-phi-findings/issues/268}{M-01}&
    \href{https://github.com/code-423n4/2024-08-phi-findings/issues/268\#issuecomment-2357330877}{Other}&
    Yes 
    \\
    \href{https://solodit.cyfrin.io/issues/h-08-player-can-mint-more-fighter-nfts-during-claim-of-rewards-by-leveraging-reentrancy-on-the-claimrewards-function-code4rena-ai-arena-ai-arena-git}{077} &
    \href{https://github.com/code-423n4/2024-02-ai-arena}{2024-02-ai-arena} &
    Players can exploit a reentrancy bug to claim extra rewards before the contract updates their NFT balance. &
    %High &
    \href{https://github.com/code-423n4/2024-02-ai-arena-findings/issues/37}{H-08}&
    \href{https://github.com/ArenaX-Labs/2024-02-ai-arena-mitigation/pull/6/files\#diff-b7b791431bf00bf243ef885bca223669bc5c7970e24202017c3736b65c62ed1f}{\#PR6}&
    Yes 
    \\
    \href{https://solodit.cyfrin.io/issues/h-01-pumps-are-not-updated-in-the-shift-and-sync-functions-allowing-oracle-manipulation-code4rena-basin-basin-git}{091} &
    \href{https://github.com/code-423n4/2023-07-basin}{2023-07-basin} &
    Users can manipulate the reported asset reserves, causing incorrect price data. & 
    %High &
    \href{https://github.com/code-423n4/2023-07-basin-findings/issues/136}{H-01}&
    \href{https://github.com/BeanstalkFarms/Basin/pull/97/files}{\#PR97}&
    Yes 
    \\
    %%%%%%%%%%%%
    \href{https://solodit.cyfrin.io/issues/h-03-wp-h0-fake-balances-can-be-created-for-not-yet-existing-erc20-tokens-which-allows-attackers-to-set-traps-to-steal-funds-from-future-users-code4rena-cally-cally-contest-git}{098} &
    \href{https://github.com/code-423n4/2022-05-cally}{2022-05-cally} &
    Fake token balances can be created for nonexistent ERC20s, enabling traps that steal funds from later users. &
    %High &
    \href{https://github.com/code-423n4/2022-05-cally-findings/issues/225}{H-03}&
    \href{https://github.com/outdoteth/cally/pull/5}{\#PR5}&
    No 
    \\
    \SetCell[c=1]{r}\textbf{Total} &  \textbf{23} && \textbf{M:15 \\ H:8} & &\textbf{Y:13 \\ N: 10} \\

\end{tblr}
\caption{Proof-of-Patch Dataset Overview}
\label{tab:proof_of_patch_dataset}
\end{table*}

%number of projects
%LoC distribution of original code 
%LOC of proposed PoC (after manual validation)
%We add the dataset config on README if needed.

\subsection{Model Selection}\label{sec:model-selection}
To evaluate \tool we select three LLMs that are appropriate for our downstream task of generating Solidity PoCs.
Also, those three models keep inference costs within a reasonable budget.

First, we select two models based on their top performance on LiveBench's agentic coding leaderboard\footnote{\href{https://livebench.ai/}{https://livebench.ai/} sorted by "agentic coding" as of 2025-10-17}: the highest-ranked closed-source model and the highest-ranked open-weight model.  
% https://livebench.ai/#/?Reasoning=a&Coding=a&Agentic+Coding=a
\textbf{\claude \cite{anthropic_claude_sonnet_4_5} (Frontier):} At the time of writing, Anthropic's \claude (2025-09-29) is currently the highest-performing commercial model for code reasoning and agentic tasks. It represents the upper bound of current capabilities (\$3/M input tokens \$15/M output tokens 1,000,000 context\footnote{\href{https://openrouter.ai/anthropic/claude-sonnet-4.5}{https://openrouter.ai/anthropic/claude-sonnet-4.5}}).  
\textbf{\glm \cite{zai_glm46} (Leading open-weight model):} Z.ai's \glm (2025-09-30), a fully open model that can be deployed locally or on cloud hardware. (\$0.50/M input tokens \$1.75/M output tokens, 202,752 context\footnote{\href{https://openrouter.ai/z-ai/glm-4.6}{https://openrouter.ai/z-ai/glm-4.6}})

Second, we select a third model based on its performance in recent related literature. \textbf{\openai \cite{openai_o3_o4mini_system_card_2025}:} OpenAI's o3 (2025-04-16) is a reasoning model used in recent exploit generation work~\cite{gervais2025ai} (\$2/M input tokens, \$8/M output tokens, 200,000 context)\footnote{\href{https://openrouter.ai/openai/o3}{https://openrouter.ai/openai/o3}}. We select o3 rather than o3-pro due to cost constraints.

\subsection{Baselines}
Three similar research projects exist for automated exploit generation for smart contracts: REX \cite{xiao2025prompt2pwn}, A1 \cite{gervais2025ai}, and Quimera\footnote{\href{https://github.com/gustavo-grieco/quimera/}{https://github.com/gustavo-grieco/quimera/}.
However, a direct quantitative comparison with them is not feasible. The implementations of REX and A1 are not publicly available, and the authors did not provide access upon request.
% we can discard quimera because 1) it has an UI interface, unfeasible to automated evaluation and reproducibility 2) we have not been able to set up the tool
% Quimera is open-source but relatively immature at the time of writing. Despite consulting the documentation and opening an issue requesting assistance, we were unable to successfully run the tool.
Quimera is open-source but relatively immature at the time of writing. Despite consulting the documentation and opening an issue requesting assistance, we were unable to successfully run the tool.}
%we opened an issue asking for help and better instructions.

Therefore, we instead choose to design two controlled baselines to measure the contribution of \tool's agentic capabilities. These baselines are chosen to represent a progression of autonomy and to test whether simpler approaches can achieve comparable results.

The two baselines, in order of increasing autonomy, are:
\begin{enumerate}
    \item \textbf{Zero-shot Prompting:} An  LLM is prompted to generate the complete PoC exploit in a single attempt, without any iterative refinement with execution feedback. This baseline represents the simplest, non-agentic LLM-based approach, isolating the raw power of the considered LLM.
    \item \textbf{Workflow Prompting:} A single LLM follows a fixed, two-phase workflow:
    (1) annotation analysis and (2) PoC generation. While the model receives execution feedback from compilation and test runs and may iteratively correct errors, the workflow structure, available actions, and phase ordering are pre-defined. The model does not perform autonomous planning, goal decomposition, or dynamic tool selection (Sec. \ref{sec:basic-tools}).
\end{enumerate}

These baselines reflect common practices in recent literature. Zero-shot Prompting represents the standard LLM querying approach widely studied before agentic advancements \cite{brown2020language}, while Workflow Prompting reflects structured, feedback-driven pipelines such as Agentless \cite{xia2024agentless}.
In comparison, \tool is agentic in that it autonomously plans, decomposes, and executes goals across multiple steps. Unlike pre-defined workflows, it can adjust its action sequence and tool usage based on intermediate results and current information. While Workflow Prompting is limited to reading the specified vulnerable file and its corresponding annotation, writing a PoC, and executing the compilation and test tools (Section~\ref{sec:domain-tools}), \tool can additionally search the broader codebase, locate symbols via grep/glob, perform partial file edits, and plan subsequent actions, as detailed in Section~\ref{sec:agentic-arch}.

The comparison of \tool with the baselines is fair as follows.
The baselines, as well as \tool, receive the same initial input: the vulnerable contract and the auditor's vulnerability annotations in natural language. They also follow identical stopping criteria, as described in Section~\ref{subsec:rq1-methodology}.

\subsection{Research Questions}
We design the following research questions to evaluate the PoC exploits generated by \tool: 
\begin{itemize}
    \item RQ1 \rqone
    \item RQ2 \rqtwo
    \item RQ3 \rqthree
\end{itemize}

\subsection{RQ1: \rqone}
\label{subsec:rq1-methodology}
We assess the capability of \tool to generate executable PoC exploits for smart contracts, based on vulnerability descriptions from security auditors. 

Per the definition of Sec. \ref{sec:definitions}, a PoC exploit corresponds to an executable test. In our experiments, they are expected to be written in the smart contract testing framework Foundry. 
Hence, we check that a generated PoC can be compiled and executed.

We define a well-formed PoC exploit as one where 1) the PoC compiles, and 2) all test assertions in the generated PoC pass.
Conversely, a failed attempt is any of the following: \begin{enumerate}
  \item \textbf{Compilation failure:} the generated PoC contains compilation errors.
  \item \textbf{No Assertion:} the test case does not include assertions, hence does not assert any wrong behavior.
  \item \textbf{Ill-formed Assertion:} an assertion in the generated PoC fails during exploit execution.
  \item \textbf{Max Cost:} the cumulative execution cost for the run exceeds a threshold of \maxcost, to keep our experimental budget under control. % similarly to previous work 
  \item \textbf{Max Tool Calls:} the total number of calls to the smart contract tools (\verb|smart-contract-compile| and \verb|smart-contract|
  \verb|-test|) exceeds a threshold ($10$).  
\end{enumerate}

We compare \tool against the baseline approaches for each of the three evaluated models. We run each PoC generation attempt once per prior agentic evaluation protocols \cite{abramovich2025enigma}.

\textbf{Metric.} For failed cases, we report their cause of origin: compilation failure (CF), no assertion (NA), ill-formed assertion (IA), max cost exceeded (MC), and max tool calls exceeded (MT).
We report the number of successful runs aggregated over all projects in \dataset. 

If multiple failure modes occur, we report only the first encountered failure. Resource-limit failures (MC and MT) take precedence, as they terminate execution before semantic validation can be completed (MC > MT > CF > IA > NA).
%For the workflow agent, which is deterministic by design, compilation failures (CF) and ill-formed assertions (IA) cannot occur in the final state, because the workflow either produces a valid executable test or terminates due to resource limits (MC/MT). 
For Zero-shot Prompting, resource limits (MC/MT) do not apply, since each run executes a single iteration.

\subsection{RQ2: \rqtwo }\label{subsec:rq2-methodology}
This RQ evaluates whether the PoC exploits generated in RQ1 demonstrate real, executable attacks against the target contracts.
Our core insight is to use the corresponding mitigation patches as evidence of logical correctness.

%main idea
% R2.5

Prior work has studied how to use exploits to assess the quality of automated smart contract patches~\cite{bobadilla2025automated}, as well as how to derive exploits from patches~\cite{brumley2008automatic}. In contrast, we use patches as a post-hoc oracle to validate the correctness of the patch-independent PoC exploits generated in RQ1.
A logically correct PoC must exercise a vulnerable execution path and therefore be mitigated by the corresponding patch.

% R3.1a
Our methodology treats the developer-provided patch as an oracle for PoC validity. This assumes that the patch is correct; in other words, the developer patch is the ground-truth for how the vulnerability should be fixed.
%R3.1a

If the developer patch is incomplete, i.e., it mitigates only a specific manifestation of a vulnerability while leaving semantically equivalent variants unaddressed, a correct PoC may still exploit the patched code. To account for this, we record a PoC as logically correct if at least one PoC assertion fails when executed on the patched code. This criterion ensures that PoC validation does not rely on patch completeness and is sound even if the patch mitigates only a subset of exploit variants. Conversely, if no assertion fails on the patched version, we conclude that the PoC does not exercise the vulnerability mitigated by the ground-truth patch.

Consequently, we validate a PoC by (1) demonstrating successful exploitation of the vulnerable code and (2) demonstrating mitigation when executed against the patch. 
Therefore, under the assumption that the patch is reliable, if the PoC is blocked by it, we can be highly confident that the PoC exposes a genuine attack path. 
Since the patches in our dataset are obtained directly from the project's development team and confirmed through manual curation, they serve as strong validation oracles for this evaluation scheme.

For each \textit{well-formed} PoC generated in RQ1, we execute it against the ground-truth mitigation patch using the same test harness and configuration. 
We define three outcomes:
\begin{enumerate}
    \item \textbf{Correct PoC:} The PoC executes, and at least one PoC assertion fails on the patched version, indicating that the patch prevents the exploit.
    \item \textbf{Incorrect PoC:} No assertion fails when the generated PoC is executed on the patched project, indicating the PoC does not exercise the vulnerability addressed by the ground-truth patch.
    \item \textbf{Inconclusive:} The patch introduces a change that breaks the compilability of the PoC exploit contract. Thus, one cannot automatically assess the validity of the exploit. In this case, we revert to manual analysis.
\end{enumerate}

This procedure is fully automated, enabling reproducible and systematic validation of all generated PoCs.
To the best of our knowledge, we are the first to devise and perform this mitigation-based validation methodology for proof-of-concept assessment.

\textbf{Metrics.} For each well-formed PoC from RQ1, we report its success when run against the patched code. For failed cases, we report their cause: incorrect (IC) or inconclusive (IN).

% Link to the canva image:
%https://www.canva.com/design/DAGzxw0lSPY/aZ7c6tfqCnPgEMc08taMmQ/edit?utm_content=DAGzxw0lSPY&utm_campaign=designshare&utm_medium=link2&utm_source=sharebutton

\subsection{RQ3: \rqthree }\label{subsec:rq3-methodology}
To assess how annotation quality affects \tool's performance, we systematically vary the descriptive detail of the original vulnerability annotations.

% what is in an annotation
We define three cumulative levels of detail:
\begin{enumerate}
    \item \textbf{High-level}: provides a high-level summary of the vulnerability's type and affected components.
    \item \textbf{Detailed}: adds code snippets and technical explanation of the root cause.
    \item \textbf{Procedural}: additionally includes step-by-step exploitation scenario in natural language.
\end{enumerate}
% building on top of each other
Each level is a strict superset of the previous one. The key discriminator between detailed and procedural is whether the annotation describes what is wrong versus how to reproduce it.
%For five cases, we remove overly detailed content at the abstract level to make sure that all annotations are at comparable levels of detail.

% missing levels
Not all annotations in our dataset contain sufficient detail to support all three levels. We restrict this analysis to the nine samples where all three versions could be constructed (see Appendix~\ref{app:annotation-levels}). For each annotation level, we run \tool with \claude using the same setup as RQ2, reporting logical correctness.

%\textbf{Metrics.}
\section{Experimental Results}
\label{sec:experimental_results}

\subsection{Experiment Settings}
We conducted our experiments on a server running Ubuntu 22.04, equipped with a 36-core Intel Core i9-10980XE CPU at 3.00 GHz and 125 GB of RAM. 
%The model temperature was set to 0 to elicit more deterministic outputs that reflect the model’s most representative learned behavior. 
The experiments were conducted on October 29, 2025. We configured the models' temperature to 0 and used a seed of $1615315$. The total experimental budget amounted to \$135 USD. % if we include an additional erroneous run it is + 170 dollars. the development budget was like 100 dollars

\subsection{RQ1 Results}
Table~\ref{tab:RQ1_result} presents the results for RQ1 evaluating the well-formedness of generated PoC exploits on the \dataset dataset. For each case, we report the success per model for \tool and the two baselines: Zero-shot and Workflow Prompting.
\SetTblrInner{rowsep=1pt, colsep=2pt}
\begin{table*}
\centering
\begin{tblr}{
    colspec ={cl ccc  ccc  ccc },
    hline{1,3,26,33} ={solid},
    vline{3,6,9}={solid},
    hline{31}={dashed},
    vline{4,5,7,8,10,11}={dashed},
    row{Z} = {font=\bfseries},
    %has POC008,015,018,020, 033, 041,042,046,048,054,070,077,091
    row{5,7,8,9,11,13,14,15,16,19,22,23,24} = {gray!30!white},
}
\SetCell[r=2]{c}{\textbf{ID}} & 
\SetCell[r=2]{c}{\textbf{Project}} & 
\SetCell[c=3]{c}{\textbf{Zero-shot}} & & &
\SetCell[c=3]{c}{\textbf{Workflow}} & & &
\SetCell[c=3]{c}{\textbf{\tool}} & & 
\\
& &
\rotatebox{80}{GLM 4.6} &
\rotatebox{80}{OpenAI o3} & 
\rotatebox{80}{\claude }&
\rotatebox{80}{GLM 4.6} &
\rotatebox{80}{OpenAI o3} & 
\rotatebox{80}{\claude }&
\rotatebox{80}{GLM 4.6} &
\rotatebox{80}{OpenAI o3} &
\rotatebox{80}{\claude }&
\\
% ---------- Rows ----------
\href{https://solodit.cyfrin.io/issues/m-01-multicall-does-not-work-as-intended-code4rena-size-size-git}{001} &
\href{https://github.com/code-423n4/2024-06-size}{2024-06-size} & 
CF & CF & CF &
MT & \cmark & \cmark &
MT & \cmark & \cmark
\\
\href{https://solodit.cyfrin.io/issues/h-04-vaultmintyieldfee-function-can-be-called-by-anyone-to-mint-vault-shares-to-any-recipient-address-code4rena-pooltogether-pooltogether-git}{003} & 
\href{https://github.com/code-423n4/2023-07-pooltogether}{2023-07-pooltogether} &
CF & IA & CF &
MT & \cmark & \cmark &
\cmark & \cmark & \cmark
\\
\href{https://solodit.cyfrin.io/issues/m-05-investors-claiming-their-maxdeposit-by-using-the-liquiditypooldeposit-will-cause-other-users-to-be-unable-to-claim-their-maxdepositmaxmint-code4rena-centrIAuge-centrIAuge-git}{008} &
\href{https://github.com/code-423n4/2023-09-centrIAuge}{2023-09-centrifuge} &
CF & CF & CF &
MT & MT & MT &
\cmark & NA & MC
\\
\href{https://solodit.cyfrin.io/issues/m-08-loss-of-funds-for-traders-due-to-accounting-error-in-royalty-calculations-code4rena-caviar-caviar-private-pools-git}{009} &
\href{https://github.com/code-423n4/2023-04-caviar}{2023-04-caviar} &
CF & CF & CF &
MT & MT & \cmark &
\cmark & MC & \cmark
\\
{015} &
\href{https://github.com/code-423n4/2023-07-pooltogether}{2023-07-pooltogether} & 
CF & CF & CF &
MT & MT & \cmark &
\cmark & \cmark & \cmark
\\
\href{https://solodit.cyfrin.io/issues/m-15-pool-tokens-can-be-stolen-via-privatepoolflashloan-function-from-previous-owner-code4rena-caviar-caviar-git}{018} &
\href{https://github.com/code-423n4/2023-04-caviar}{2023-04-caviar} &
CF & CF & IA &
MT & MT & MT &
\cmark & \cmark & MC
\\
\href{https://solodit.cyfrin.io/issues/m-3-share-price-inflation-by-first-lp-er-enabling-dos-attacks-on-subsequent-buyshares-with-up-to-1001x-the-attacking-cost-sherlock-dodo-gsp-git}{020} & 
\href{https://github.com/sherlock-audit/2023-12-dodo-gsp}{2023-12-dodo-gsp} &
CF & CF & CF &
MT & \cmark & \cmark &
\cmark & \cmark & \cmark
\\
\href{https://solodit.cyfrin.io/issues/m-06-denial-of-service-contract-owner-could-block-users-from-withdrawing-their-strike-code4rena-putty-putty-contest-git}{032} &
\href{https://github.com/code-423n4/2022-06-putty}{2022-06-putty} &
CF & CF & CF &
MT & MT & MT &
\cmark & \cmark & MC
\\
\href{https://solodit.cyfrin.io/issues/m-03-flash-loan-fee-is-incorrect-in-private-pool-contract-code4rena-caviar-caviar-private-pools-git}{033} &
\href{https://github.com/code-423n4/2023-04-caviar}{2023-04-caviar} &
CF & CF & CF &
MT & \cmark & MT &
\cmark & \cmark & \cmark
\\
\href{https://solodit.cyfrin.io/issues/h-2-m-1-sherlock-axis-finance-git}{039} &
\href{https://github.com/sherlock-audit/2024-03-axis-finance}{2024-03-axis-finance} &
CF & IA & CF &
MT & MT & \cmark &
MT & MC & \cmark
\\
\href{https://solodit.cyfrin.io/issues/h-1-malicious-user-can-overtake-a-prefunded-auction-and-steal-the-deposited-funds-sherlock-axis-finance-git}{041} &
\href{https://github.com/sherlock-audit/2024-03-axis-finance}{2024-03-axis-finance} &
CF & CF & CF &
MT & \cmark & MT &
IA & \cmark & \cmark
\\
\href{https://solodit.cyfrin.io/issues/m-2-utilization-rate-multiplier-will-not-shIAt-IA-oracle-is-consulted-frequently-sherlock-cap-git}{042} &
\href{https://github.com/sherlock-audit/2025-07-cap}{2025-07-cap} &
CF & CF & CF &
MT & MT & MT &
\cmark & \cmark & MC
\\
\href{https://solodit.cyfrin.io/issues/m-03-zero-token-transfer-can-cause-a-potential-dos-in-cvxstaker-code4rena-xeth-xeth-versus-contest-git}{046} &
\href{https://github.com/code-423n4/2023-05-xeth}{2023-05-xeth} &
CF & CF & CF &
MT & \cmark & \cmark &
\cmark & \cmark & \cmark
\\
\href{https://solodit.cyfrin.io/issues/h-01-royalty-receiver-can-drain-a-private-pool-code4rena-caviar-caviar-private-pools-git}{048} &
\href{https://github.com/code-423n4/2023-04-caviar}{2023-04-caviar} &
CF & CF & IA &
MT & MT & MT &
MT & MC & MC
\\
\href{https://solodit.cyfrin.io/issues/m-2-lender-is-able-to-steal-borrowers-collateral-by-calling-rollloan-with-unfavourable-terms-on-behalf-of-the-borrower-sherlock-cooler-update-git}{049} &
\href{https://github.com/sherlock-audit/2023-08-cooler}{2023-08-cooler} &
CF & IA & CF &
MT & MT & MT &
\cmark & \cmark & \cmark
\\
\href{https://solodit.cyfrin.io/issues/m-04-you-can-deposit-really-small-amount-for-other-users-to-dos-them-code4rena-centrIAuge-centrIAuge-git}{051} &
\href{https://github.com/code-423n4/2023-09-centrIAuge}{2023-09-centrifuge} &
IA & CF & CF &
MT & \cmark & \cmark &
\cmark & \cmark & MC
\\
\href{https://solodit.cyfrin.io/issues/h-01-no-revert-on-transfer-erc20-tokens-can-be-drained-code4rena-cally-cally-contest-git}{054} &
\href{https://github.com/code-423n4/2022-05-cally}{2022-05-cally} &
CF & \cmark & CF &
MT & \cmark & MT &
\cmark & \cmark & \cmark
\\
\href{https://solodit.cyfrin.io/issues/m-05-fillorder-and-exercise-may-lock-ether-sent-to-the-contract-forever-code4rena-putty-putty-git}{058} &
\href{https://github.com/code-423n4/2022-06-putty}{2022-06-putty} &
CF & \cmark & CF &
MT & \cmark & MT &
MT & \cmark & \cmark
\\
\href{https://solodit.cyfrin.io/issues/h-02-protocol-mints-less-rseth-on-deposit-than-intended-code4rena-kelp-dao-kelp-dao-git}{066} &
\href{https://github.com/code-423n4/2023-11-kelp}{2023-11-kelp} &
CF & CF & CF &
MT & \cmark & \cmark &
\cmark & \cmark & MC
\\
\href{https://solodit.cyfrin.io/issues/m-01-contract-phinft1155-cant-be-paused-code4rena-phi-phi-git}{070} &
\href{https://github.com/code-423n4/2024-08-ph}{2024-08-ph} &
CF & CF & CF &
MT & MT & MT &
\cmark & \cmark & \cmark
\\
\href{https://solodit.cyfrin.io/issues/h-08-player-can-mint-more-fighter-nfts-during-claim-of-rewards-by-leveraging-reentrancy-on-the-claimrewards-function-code4rena-ai-arena-ai-arena-git}{077} &
\href{https://github.com/code-423n4/2024-02-ai-arena}{2024-02-ai-arena} &
CF & \cmark & \cmark &
MT & \cmark & MT &
MT & \cmark & MC
\\
\href{https://solodit.cyfrin.io/issues/h-01-pumps-are-not-updated-in-the-shIAt-and-sync-functions-allowing-oracle-manipulation-code4rena-basin-basin-git}{091} &
\href{https://github.com/code-423n4/2023-07-basin}{2023-07-basin} &
CF & CF & CF &
MT & \cmark & MT &
MT & \cmark & \cmark
\\
\href{https://solodit.cyfrin.io/issues/h-03-wp-h0-fake-balances-can-be-created-for-not-yet-existing-erc20-tokens-which-allows-attackers-to-set-traps-to-steal-funds-from-future-users-code4rena-cally-cally-contest-git}{098} &
\href{https://github.com/code-423n4/2022-05-cally}{2022-05-cally} &
CF & CF & CF &
MT & \cmark & MT &
\cmark & \cmark & \cmark
\\

% ---------- Totals ----------
\SetCell[c=2]{r}\textbf{\#Compilation Failure (CF)} && 22 & 17 & 20 & \SetCell{bg=gray!20} & \SetCell{bg=gray!20}& \SetCell{bg=gray!20} & 0 & 0 & 0\\
\SetCell[c=2]{r}\textbf{\#No Assertion (NA)} && 0 & 0 & 0 & 0 & 0 & 0 & 0 & 1 & 0\\
\SetCell[c=2]{r}\textbf{\#Ill-formed Assertion (IA)} && 1 & 3 & 2 & \SetCell{bg=gray!20} & \SetCell{bg=gray!20} & \SetCell{bg=gray!20} & 1 & 0 & 0\\
\SetCell[c=2]{r}\textbf{\#Max Cost (MC)} && \SetCell{bg=gray!20}& \SetCell{bg=gray!20}& \SetCell{bg=gray!20}& 0 & 0 & 0 & 0 & 3 & 8\\
\SetCell[c=2]{r}\textbf{\#Max Tool Calls (MT)} && \SetCell{bg=gray!20}& \SetCell{bg=gray!20}& \SetCell{bg=gray!20}& 23 & 10 & 14 & 6 & 0 & 0\\
\SetCell[c=2]{r}\textbf{\#Well-formed (\cmark)} && 0 & 3 & 1 & 0 & 13 & 9 & 16 & 19 & 15\\
    \end{tblr} 
\centering
\caption{RQ1 Overview: well-formedness of generated PoCs. Gray cells indicate inapplicable categories as described in Section~\ref{subsec:rq1-methodology}.}
\label{tab:RQ1_result}
\end{table*}

% finding 1
\textbf{Zero-shot Prompting.}The simplest baseline struggles significantly to generate useful PoCs.
Out of 69 PoC exploits across 23 projects and 3 models, only 4 are well-formed. %considered valid according to our automated validation criteria. 
The 4 successfully generated cases are distributed across 2 models, \openai and \claude, and 3 projects: \#054, \#058, and \#077. It is noteworthy that, among these, only \#077 has an identified ground-truth PoC; therefore, data leakage is unlikely to explain the results for \#054 and \#058. %cannot be attributed to data leakage.
The most common error for the Zero-shot Prompting baseline is `Compilation Failure' (CF), indicating that the model struggles with syntactical and semantic issues. % and Ill-formed Assertion (IA).

% finding 2
\textbf{Workflow Prompting.} The Workflow baseline significantly improves the number of well-formed PoC cases, generating 22 well-formed exploits, 13 by  \openai and 9 by \claude.
Notably, GLM did not produce a single successful PoC for the workflow baseline, consistently reaching max tool call stop.

\textbf{\tool.} Finally, \tool's configuration is the most successful, producing a total of 50 well-formed PoC exploits, distributed among GLM, OpenAI, and Claude with 16, 19, and 15 cases, respectively.
The most common error causes for failed PoCs are as follows: \glm: max tool calls: \openai: max cost (3 out of 4 failed PoCs); \claude: max cost (8 cases). 
In one iteration, \openai terminated prematurely before generating a PoC; we classify this case as \emph{No Assertion (NA)}.
Overall, \tool's agentic approach demonstrates a clear advantage over Zero-shot and Workflow Prompting.   

% finding 4: compilation is solved
% \todo{elaborate on the remarkable fact that there is no compilation failure for workflow and prompting}

To better understand the varying performance of our approaches, we examine two representative case studies that illustrate why some baselines fail to generate executable PoCs, while the workflow and \tool\ configurations achieve significant improvements.

\subsubsection{From Compilation Errors to Logical Misconfigurations (\#041)}
%\subsubsection{Case study of 2024-03-axis-finance (\#041)}
Axis is a protocol for on-chain auctions. This vulnerability allows a malicious user to overtake a prefunded auction and steal the deposited funds. % "src/bases/Auctioneer.sol",
All models in the Zero-shot Prompting baseline fail to compile. \openai errors with a syntax issue: it creates an invalid hexadecimal literal when constructing an attacker address (see Listing~\ref{lst:041_syntax}).

\begin{listing}\begin{minted}[fontsize=\scriptsize]{bash}
$ forge test compile 
Compiler run failed:
Error (8936): Identifier-start is not allowed at end of a number.
  --> test/exploit/ExploitTest.t.sol:91:41:
   |
91 |     address internal attacker = address(0xEvil);  // malicious actor
   |                                         ^^^

Error: Compilation failed
\end{minted}
\caption{Prompting with \openai, generates a PoC with compilation error due to invalid hexadecimal literal.}
\label{lst:041_syntax}  
\end{listing}
Sonnet and GLM fail during semantic analysis for the same class of errors. Both attempt to reuse a base module from the vulnerable contract but provide a mock implementation that overrides several functions with incompatible signatures (mismatched visibility and return types).  % (e.g., \texttt{import \{Module, Keycode, Veecode\} from "../../src/modules/Modules.sol";}),
Because the Zero-shot approach cannot thoroughly explore the codebase, the models often infer missing details and redeclare functions, resulting in compilation errors related to overrides. 
% is interesting?? GLM produces a file with 363 lines, compared to 177 and 212 for o3 and Claude, respectively.

Under the workflow approach, o3 produces a well-formed PoC on its first attempt: a concise 93-line file. To avoid the override errors, it omits the module import and reimplements the vulnerable contract in a standalone \texttt{BuggyAuctioneer} harness. Unfortunately, as we will see in RQ2, the well-formed exploit never exercises the vulnerable contract.

% Sonnet and o3 do attempt to reuse the real module by to mocking it via overrides. Because they can iterate against compiler feedback, Sonnet is able to fix some issues but ultimately fails on other semantic problems (for example, undeclared identifiers). GLM does not resolve the override-induced errors in these trials,m resulting in max tool call abortion.

\begin{listing}\begin{minted}[fontsize=\scriptsize]{bash}
$ forge test --match-path test/exploit/ExploitTest.t.sol 
[⠊] Compiling...
[⠃] Compiling 1 files with Solc 0.8.1
Ran 2 tests for test/exploit/ExploitTest.t.sol:ExploitTest
[PASS] test_basicPrefundedAuction() (gas: 230986)
[FAIL] test_lotIdInitializationVulnerability_Exploit() (gas: 410455)
Suite result: FAILED. 1 passed; 1 failed; 0 skipped; finished in 6.85ms

Ran 1 test suite in 16.08ms (6.85ms CPU time): 1 tests passed, 1 failed

Failing tests:
Encountered 1 failing test in test/exploit/ExploitTest.t.sol:ExploitTest
[FAIL: Auction house should have no remaining tokens: 
    100000000000000000000 != 0] 
        test_lotIdInitializationVulnerability_Exploit() (gas: 410455)

Encountered a total of 1 failing tests, 1 tests succeeded
\end{minted}
\caption{\tool:GLM failing assertion (\#041): PoC prefunds the attacker, making the original victim funds untouched.}
\label{lst:rq1_glm_assertion}  
\end{listing}

% ill formed assertion
\emph{Ill-formed assertion.} Finally, \tool:GLM is the only \tool configuration that triggers an Ill-formed Assertion (IA). The PoC compiles on its first attempt but misconfigures the test. Instead of creating a non-prefunded auction (which would leave the victim’s deposit in the vulnerable slot), the exploit prefunds the attacker. As a result, the attacker’s own deposit overwrites slot 0 and, when the lot is later cancelled, the attacker reclaims their self-funded tokens rather than the victim’s funds. The main assertions of (1) a drained auction house and (2) the victim's loss of funds, therefore, never succeed.

%These results indicate that the LLM models can reason about reentrancy bugs, while errors in achieving it practically can be attributed to stalling in complex setups.

\begin{answerbox}{Answer to RQ1 } \textbf{\rqone}
Zero-shot and Workflow baselines struggle to produce functional smart contract exploit PoCs.  
Zero-shot generation is insufficient, and whereas the Workflow approach benefits from execution feedback, it remains limited by restricted codebase visibility.  
In contrast, \tool’s agentic architecture actively explores the project structure and leverages diagnostic feedback, achieving strong success in producing compilable, executable PoCs with assertions.

%reliable PoC generation in smart contract analysis.
%Our results highlight the need for improved tooling: Even though AI has shown impressive capabilities in code and test case generation for mainstream languages \todo{ add citation}, this is not the case for less mature languages like Solidity. Smart contract developers urgently require more reliable AI tools to support their workflows.
\end{answerbox}

%%%%%%%%%%%%%%%%%%%%%%%%%%%%%%%%%%%%%%%%%%%%%%%%%%%%%%%%%%%%%%%%%%
\subsection{RQ2 Results} \label{sec:rq2-results}
In RQ2, we assess which of the well-formed PoCs (RQ1) are also logically correct, that is, they contain a true exploit.
Recall that, according to our methodology, a correct PoC is prevented by the ground-truth mitigation patch.
Table~\ref{tab:RQ2_result} presents the result. Overall, \tool demonstrates the highest number of logically correct PoC exploits.

\textbf{Zero-shot Prompting.} Out of the 4 well-formed PoCs from RQ1, 3 are found correct by our automated validation criteria, and one is inconclusive. Of the correct cases, 2 are from \openai and one from \claude. 

\textbf{Workflow Prompting.}
The workflow baseline successfully produces logically correct exploits for 11 out of the 22 well-formed PoCs from RQ1. 
% More granulary, 8 are generated by \openai and 3 by \claude. 
%Among the incorrect PoCs, \openai accounts for 4 and \claude for 5.  
The Workflow configuration with \glm fails to generate any correct PoC.

\textbf{\tool.}
Among the well-formed PoCs evaluated in RQ2, \tool{} produces 14 logically correct exploits with the \openai configuration, 11 with \claude, and 7 with \glm, with a total of 32 correct exploits.
In terms of error ratio, \claude achieves the highest success rate at 73\%(11/15) correct PoCs, followed by \openai 68\%(13/19) and \glm 43\%(7/16).
Overall, \tool remains the most effective approach across all variants.

%Significantly, for #54,49,42 all PoC exploits are deemed incorrect according to our automated evaluation. 
%Only two findings are deemed inconclusive. 

To illustrate the practical strengths and limitations of our method, we present case studies analyzing the factors behind successful and failed PoC generation, as well as the causes of inconclusive patches.

\SetTblrInner{rowsep=1pt, colsep=2pt}

\begin{table*}
\centering
\begin{tblr}{
    colspec ={cl ccc  ccc  ccc },
    hline{1,3,26} ={solid},
    vline{3,6,9,10}={solid},
    hline{27}={solid},
    vline{4,5,7,8,10,11}={dashed},
      %has POC008,015,018,020, 033, 041,042,046,048,054,070,077,091
    row{5,7,8,9,11,13,14,15,16,19,22,23,24} = {gray!30!white},
}
\SetCell[r=2]{c}{\textbf{ID}} & 
\SetCell[r=2]{c}{\textbf{Project}} & 
\SetCell[c=3]{c}{\textbf{Zero-shot}} & & &
\SetCell[c=3]{c}{\textbf{Workflow}} & & & 
\SetCell[c=3]{c}{\textbf{\tool}} & & 
\\

&&
\rotatebox{80}{GLM 4.6} &
\rotatebox{80}{OpenAI o3} & 
\rotatebox{80}{\claude }&
\rotatebox{80}{GLM 4.6} &
\rotatebox{80}{OpenAI o3} &
\rotatebox{80}{\claude }&
\rotatebox{80}{GLM 4.6} &
\rotatebox{80}{OpenAI o3} &
\rotatebox{80}{\claude }
\\

\href{https://solodit.cyfrin.io/issues/m-01-multicall-does-not-work-as-intended-code4rena-size-size-git}{001} &
\href{https://github.com/code-423n4/2024-06-size}{2024-06-size} & 
\textemdash & \textemdash & \textemdash & \textemdash & IC & IC & \textemdash & \success & \success
\\

\href{https://solodit.cyfrin.io/issues/h-04-vaultmintyieldfee-function-can-be-called-by-anyone-to-mint-vault-shares-to-any-recipient-address-code4rena-pooltogether-pooltogether-git}{003} & 
\href{https://github.com/code-423n4/2023-07-pooltogether}{2023-07-pooltogether} &
\textemdash & \textemdash & \textemdash & \textemdash & IN & IN & IN & IN & IN
\\

\href{https://solodit.cyfrin.io/issues/m-05-investors-claiming-their-maxdeposit-by-using-the-liquiditypooldeposit-will-cause-other-users-to-be-unable-to-claim-their-maxdepositmaxmint-code4rena-centrifuge-centrifuge-git}{008} &
\href{https://github.com/code-423n4/2023-09-centrifuge}{2023-09-centrifuge} &
\textemdash & \textemdash & \textemdash & \textemdash & \textemdash & \textemdash & IC & \textemdash & \textemdash
\\

\href{https://solodit.cyfrin.io/issues/m-08-loss-of-funds-for-traders-due-to-accounting-error-in-royalty-calculations-code4rena-caviar-caviar-private-pools-git}{009} &
\href{https://github.com/code-423n4/2023-04-caviar}{2023-04-caviar} &
\textemdash & \textemdash & \textemdash & \textemdash & \textemdash & IC & \success & \textemdash & \success
\\

\href{https://solodit.cyfrin.io/issues/m-02-unintended-or-malicious-use-of-prize-winners-hooks-code4rena-pooltogether-pooltogether-git}{015} &
\href{https://github.com/code-423n4/2023-07-pooltogether}{2023-07-pooltogether} & 
\textemdash & \textemdash & \textemdash & \textemdash & \textemdash & IN & IN & IN & IN
\\

\href{https://solodit.cyfrin.io/issues/m-15-pool-tokens-can-be-stolen-via-privatepoolflashloan-function-from-previous-owner-code4rena-caviar-caviar-git}{018} &
\href{https://github.com/code-423n4/2023-04-caviar}{2023-04-caviar} &
\textemdash & \textemdash & \textemdash & \textemdash & \textemdash & \textemdash & \success & \success & \textemdash
\\

\href{https://solodit.cyfrin.io/issues/m-3-share-price-inflation-by-first-lp-er-enabling-dos-attacks-on-subsequent-buyshares-with-up-to-1001x-the-attacking-cost-sherlock-dodo-gsp-git}{020} & 
\href{https://github.com/sherlock-audit/2023-12-dodo-gsp}{2023-12-dodo-gsp} &
\textemdash & \textemdash & \textemdash & \textemdash & IC & \success & IC & \success & \success
\\

\href{https://solodit.cyfrin.io/issues/m-06-denial-of-service-contract-owner-could-block-users-from-withdrawing-their-strike-code4rena-putty-putty-contest-git}{032} &
\href{https://github.com/code-423n4/2022-06-putty}{2022-06-putty} &
\textemdash & \textemdash & \textemdash & \textemdash & \textemdash & \textemdash & \success & \success & \textemdash
\\

\href{https://solodit.cyfrin.io/issues/m-03-flash-loan-fee-is-incorrect-in-private-pool-contract-code4rena-caviar-caviar-private-pools-git}{033} &
\href{https://github.com/code-423n4/2023-04-caviar}{2023-04-caviar} &
\textemdash & \textemdash & \textemdash & \textemdash & \success & \textemdash & \success & \success & \success
\\

\href{https://solodit.cyfrin.io/issues/h-2-m-1-sherlock-axis-finance-git}{039} &
\href{https://github.com/sherlock-audit/2024-03-axis-finance}{2024-03-axis-finance} &
\textemdash & \textemdash & \textemdash & \textemdash & \textemdash & IC & \textemdash & \textemdash & \success
\\

\href{https://solodit.cyfrin.io/issues/h-1-malicious-user-can-overtake-a-prefunded-auction-and-steal-the-deposited-funds-sherlock-axis-finance-git}{041} &
\href{https://github.com/sherlock-audit/2024-03-axis-finance}{2024-03-axis-finance} &
\textemdash & \textemdash & \textemdash & \textemdash & IC & \textemdash & \textemdash & \success & \success
\\

\href{https://solodit.cyfrin.io/issues/m-2-utilization-rate-multiplier-will-not-shift-if-oracle-is-consulted-frequently-sherlock-cap-git}{042} &
\href{https://github.com/sherlock-audit/2025-07-cap}{2025-07-cap} &
\textemdash & \textemdash & \textemdash & \textemdash & \textemdash & \textemdash & IC & IC & \textemdash
\\

\href{https://solodit.cyfrin.io/issues/m-03-zero-token-transfer-can-cause-a-potential-dos-in-cvxstaker-code4rena-xeth-xeth-versus-contest-git}{046} &
\href{https://github.com/code-423n4/2023-05-xeth}{2023-05-xeth} &
\textemdash & \textemdash & \textemdash & \textemdash & \success & \success & \success & \success & \success
\\

\href{https://solodit.cyfrin.io/issues/h-01-royalty-receiver-can-drain-a-private-pool-code4rena-caviar-caviar-private-pools-git}{048} &
\href{https://github.com/code-423n4/2023-04-caviar}{2023-04-caviar} &
\textemdash & \textemdash & \textemdash & \textemdash & \textemdash & \textemdash & \textemdash & \textemdash & \textemdash
\\

\href{https://solodit.cyfrin.io/issues/m-2-lender-is-able-to-steal-borrowers-collateral-by-calling-rollloan-with-unfavourable-terms-on-behalf-of-the-borrower-sherlock-cooler-update-git}{049} &
\href{https://github.com/sherlock-audit/2023-08-cooler}{2023-08-cooler} &
\textemdash & \textemdash & \textemdash & \textemdash & \textemdash & \textemdash & IC & \success & IC
\\

\href{https://solodit.cyfrin.io/issues/m-04-you-can-deposit-really-small-amount-for-other-users-to-dos-them-code4rena-centrifuge-centrifuge-git}{051} &
\href{https://github.com/code-423n4/2023-09-centrifuge}{2023-09-centrifuge} &
\textemdash & \textemdash & \textemdash & \textemdash & \success & \success & IC & \success & \textemdash
\\

\href{https://solodit.cyfrin.io/issues/h-01-no-revert-on-transfer-erc20-tokens-can-be-drained-code4rena-cally-cally-contest-git}{054} &
\href{https://github.com/code-423n4/2022-05-cally}{2022-05-cally} &
\textemdash & IC & \textemdash & \textemdash & \success & \textemdash & IC & IC & IC
\\

\href{https://solodit.cyfrin.io/issues/m-05-fillorder-and-exercise-may-lock-ether-sent-to-the-contract-forever-code4rena-putty-putty-git}{058} &
\href{https://github.com/code-423n4/2022-06-putty}{2022-06-putty} &
\textemdash & \success & \textemdash & \textemdash & \success & \textemdash & \textemdash & \success & \success
\\

\href{https://solodit.cyfrin.io/issues/h-02-protocol-mints-less-rseth-on-deposit-than-intended-code4rena-kelp-dao-kelp-dao-git}{066} &
\href{https://github.com/code-423n4/2023-11-kelp}{2023-11-kelp} &
\textemdash & \textemdash & \textemdash & \textemdash & \success & IC & IC & \success & \textemdash
\\

\href{https://solodit.cyfrin.io/issues/m-01-contract-phinft1155-cant-be-paused-code4rena-phi-phi-git}{070} &
\href{https://github.com/code-423n4/2024-08-ph}{2024-08-ph} &
\textemdash & \textemdash & \textemdash & \textemdash & \textemdash & \textemdash & \success & \success & \success
\\

\href{https://solodit.cyfrin.io/issues/h-08-player-can-mint-more-fighter-nfts-during-claim-of-rewards-by-leveraging-reentrancy-on-the-claimrewards-function-code4rena-ai-arena-ai-arena-git}{077} &
\href{https://github.com/code-423n4/2024-02-ai-arena}{2024-02-ai-arena} &
\textemdash & \success & \success  & \textemdash & \success & \textemdash & \textemdash & \success & \textemdash
\\

\href{https://solodit.cyfrin.io/issues/h-01-pumps-are-not-updated-in-the-shift-and-sync-functions-allowing-oracle-manipulation-code4rena-basin-basin-git}{091} &
\href{https://github.com/code-423n4/2023-07-basin}{2023-07-basin} &
\textemdash & \textemdash & \textemdash & \textemdash & IC & \textemdash & \textemdash & IC & \success
\\

\href{https://solodit.cyfrin.io/issues/h-03-wp-h0-fake-balances-can-be-created-for-not-yet-existing-erc20-tokens-which-allows-attackers-to-set-traps-to-steal-funds-from-future-users-code4rena-cally-cally-contest-git}{098} &
\href{https://github.com/code-423n4/2022-05-cally}{2022-05-cally} &
\textemdash & \textemdash & \textemdash & \textemdash & \success & \textemdash & \success & \success & \success
\\
\SetCell[c=2]{r}\textbf{\# Evaluated} && 0 & 3 & 1 & 0 & 13 & 9 & 16 & 19 & 15 \\

\SetCell[c=2]{r}\textbf{\# Incorrect (IC)}       && 0 & 1 & 0& 0 & 4 & 4 & 7 & 3 & 2  \\
\SetCell[c=2]{r}\textbf{\# Inconclusive (IN)}  && 0 & 0 & 0 & 0 & 1 & 2 & 2 & 2 & 2 \\
\SetCell[c=2]{r}\textbf{\# Correct (\success)} && 0 & 2 & 1& 0 & 8 & 3 & 7 & 14 & 11  \\
\end{tblr}
\centering
\caption{RQ2 Overview: Logical correctness of generated PoCs. \textemdash indicates PoCs failing the RQ1 evaluation.}
\label{tab:RQ2_result}
\end{table*}

\subsubsection{Inconclusive Patches (\#003 and \#015)}%\vivi{this is really manual correctness analysis. Consider moving under that subheading.}

All PoC exploits generated for \#003 and \#015 from the project \textit{2023-07-pool-together} are classified as inconclusive. This category indicates that execution fails after patch application, preventing us from automatically determining whether the patch logically resolves the vulnerability.

For case \#003, the upstream patch changes the parameter list of \texttt{mintYieldFee}. The PoC, which invokes the original signature, fails to compile against the patched code because its call sites and expected types no longer match.  
Similarly, for \#015, the patch modifies the function  \texttt{beforeClaimPrize} by adding two parameters. This again leads to consistent compilation failures when applying the PoC to the patched code. We manually verified all \tool exploits in this category and confirmed their correctness.

\subsubsection{Iterative Exploit Construction (\#091)}
%\subsubsection{Correct PoC for \#091 with \tool \& \claude}
\label{sec:trajectory}
In \#091, \tool is tasked with proving a price-oracle manipulation. 
%n which the \texttt{Well} contract's \texttt{shift()} and \texttt{sync()} entry points bypass \texttt{\_updatePumps()}. 
This vulnerability enables reserves to be manipulated within a single block, providing downstream protocols with poisoned prices that can lead to mispriced swaps and collateral distortions.
\tool can generate a correct PoC for this vulnerability.
Listing~\ref{fig:091-trajectory} summarizes the trajectory.
It shows \tool's iterative process of code generation, error diagnosis, and logic refinement. In the following, we examine each phase in detail.

\begin{figure*}[ht]
  \centering
  \small
  
  \captionsetup{type=lstlisting}
  \begin{tcolorbox}[
      colback=blue!4,             % very light blue background
      colframe=blue!50!black,     % muted professional blue border
      width=\textwidth,
      arc=2mm,
      boxrule=0.6pt,              % slightly softer border
      left=6pt,
      right=6pt,
      top=6pt,
      bottom=6pt]

  \textbf{PoC Trajectory: Claude Sonnet 4.5 (\#091)}\\[4pt]

  \begin{tabularx}{\textwidth}
    {@{}>{\raggedleft\arraybackslash\footnotesize}p{0.7cm}
       p{1.5cm}
       @{\hspace{0.6em}} X@{}}

  [seq.~03]  & \textsc{Assistant}: &
  “I'll analyze the vulnerability and create a PoC test. Let me start by examining the vulnerable contract and the vulnerability description.” \\[3pt]

  [seq.~29] & \textsc{Tool}: &
  \texttt{forge test} aborts with Error 7576 — \texttt{console.log} used without importing \texttt{forge-std/Test.sol}. \\[3pt]

  [seq.~42] & \textsc{Assistant}: &
  “I see, the IPump interface only has \texttt{update}, not \texttt{read}. Let me fix the TrackingPump.” \\[3pt]

  [seq.~49] & \textsc{Tool}: &
  Test suite run: two passes, \texttt{test\_exploit\_sync\_oracle\_manipulation} still fails because the oracle assertion expects post-swap reserves. \\[3pt]

  [seq.~56] & \textsc{Assistant}: &
  “The issue is in line 151 — the pump receives reserves \emph{before} the swap adds to them. I’ll adjust the assertion to compare with \texttt{reservesAfterSync}.” \\[3pt]

  [seq.~63] & \textsc{Tool}: &
  Final \texttt{forge test} succeeds — all three tests now pass, validating the oracle-manipulation PoC.

  \end{tabularx}

  \end{tcolorbox}
  \caption{Extracted moments from the Basin \#091 oracle-manipulation trajectory: reconnaissance, incremental debugging, and the final passing PoC.}
  \label{fig:091-trajectory}
\end{figure*}

\begin{figure}[b]
\input{listings/rq2/rq2_091} 
\end{figure}

\textbf{Reconnaissance (rounds~1-~21).}
\tool begins by reading the linked code segments of \texttt{Well.sol} described in the annotation, the Foundry configuration, TestHelper utilities, and existing pump mocks. This investigation tells \tool how the Basin test harness provisions wells, tokens, and pumps, and clarifies the attack surface described in the annotation.

\textbf{Initial Exploit Writing (round~26).}
With that context, \tool authors \texttt{ExploitTest.t.sol}. The initial version introduces a custom \emph{TrackingPump} that increments the \texttt{updateCount} and caches the latest reserve vector received through the \texttt{update} method. Three exploit tests are created: two attacker flows that exploit \texttt{shift()} and \texttt{sync()} respectively, and a control swap that documents the expected pre-update behaviour of \texttt{swapFrom()}.

\textbf{Compilation Debugging (rounds~27-~46).} 
Compilation proceeds in iterative passes. The first attempt fails because the suite uses
\texttt{console.log} without importing \texttt{forge-std/Test.sol}; the agent adds the
import and retries. The second attempt fails after the mock tries to override a nonexistent
\texttt{IPump.read}; consulting the interface prompts the agent to drop the override, and the
third compile succeeds.

\textbf{Test Assertion Refinement (rounds~47–~58).}
The first test run exposes a logical error: the \texttt{sync} compares the TrackingPump’s reading against post-swap reserves rather than the manipulated snapshot captured before execution. \tool rereads the test, updates the expectation to target \texttt{reservesAfterSync}, and verifies in the subsequent test run that both exploit tests now capture the poisoned oracle state while the control swap passes.
When this PoC is run against the developer-written mitigation, it fails, soundly demonstrating that it is exercising the vulnerability.
Listing~\ref{lst:rq2_091} presents the core vulnerability and the corresponding exploit approach.

\subsubsection{Manual Correctness Analysis}

\paragraph{\textbf{Reports with existing PoCs.}}\phantomsection\label{par:correctness-ground-truth-pocs}
Among the 23 samples in our dataset, 13 have existing PoCs written by the reporting auditor, and we consider 6/13 that have a corresponding PoCo PoC. For these six cases, we compare the PoCs generated by \tool with \claude against the original PoC to assess semantic equivalence, i.e., whether they exploit the same vulnerability. Each case was independently reviewed by two authors, with disagreements resolved by a third. 

%For each test that fails when executed against the patched version of the codebase, we identify the failing exploit assertion. %ure condition and extract the corresponding exploit assertion.

Table~\ref{tab:manual_rq2} reports all tests per PoC, passing and failing test cases, and the assertions that verify the exploit.
PoC test failures arise in two situations: the exploit assertion no longer holds (e.g., \texttt{expectRevert()} no longer triggered), or the exploit operation itself fails (e.g., a privileged call or transfer no longer succeeds).
Out of 16 PoC tests, 14 successfully fail against the developer patch: this means that the exploit itself fails under the assumption that the patch prevents the vulnerable behavior. Two tests still pass; these intentionally verify normal protocol behavior and are therefore expected to succeed after patching. In those two cases, the agent overachieves beyond PoC generation.

Overall, all 6/6 analyzed cases were deemed correct PoCs, as seen in Table~\ref{tab:manual_rq2}. 
We note that several generated PoCs contain multiple tests. 
For each test that fails when executed against the patched version of the codebase, we identify the failing exploit condition. %ure condition and extract the corresponding exploit assertion.

% the generated PoC are better than the developer ones
Beyond correctness, our analysis shows that the generated PoCs sometimes improve coverage compared to the developer-provided ones. The original PoCs often show a single exploit instance, while \tool generates multiple test variants for the same bug. This is reflected in Table~\ref{tab:manual_rq2} for \href{https://solodit.cyfrin.io/issues/m-03-zero-token-transfer-can-cause-a-potential-dos-in-cvxstaker-code4rena-xeth-xeth-versus-contest-git}{\#046}: \tool produces three tests that differ in setup (reward pool transfer vs direct minting) and time (including a 30-day warp with additional rewards), while the developer PoC exercises a single exploit case.

\begin{table*}
\scriptsize
\centering
\begin{tblr}
{
    colspec ={l l l l l l},
    hline{1,2,Z} ={solid},
    %vline{3,6,9,10}={solid},
    %hline{3,8,10,13,15}={dashed},
    row{2,8,9,13,14} = {gray!30!white}
    %vline{4,5,7,8,10,11}={dashed}
}
\textbf{ID} & \textbf{Project} & \textbf{\#Tests} & \textbf{Verdict} & \textbf{PoC Test Outcome (\xmark~ / \cmark~ ~on patch)} & \textbf{Exploit Assertion/Reverted Call} \\

% \href{https://solodit.cyfrin.io/issues/m-3-share-price-inflation-by-first-lp-er-enabling-dos-attacks-on-subsequent-buyshares-with-up-to-1001x-the-attacking-cost-sherlock-dodo-gsp-git}{\textbf{020}} &\href{https://github.com/sherlock-audit/2023-12-dodo-gsp}{\textbf{2023-12-dodo-gsp}} &1 & Yes
% & \xmark~ test\_buyShares\_lowLiquidity & \texttt{gsp.buyShares(attacker)} \\

\mbox{\textbf{\href{https://solodit.cyfrin.io/issues/m-3-share-price-inflation-by-first-lp-er-enabling-dos-attacks-on-subsequent-buyshares-with-up-to-1001x-the-attacking-cost-sherlock-dodo-gsp-git}{020}}}
&
\mbox{\textbf{\href{https://github.com/sherlock-audit/2023-12-dodo-gsp}{2023-12-dodo-gsp}}}
& 1 & Correct
& \xmark~ test\_buyShares\_lowLiquidity
& \texttt{gsp.buyShares(attacker)} \\

\SetCell[r=5]{l}{\textbf{\href{https://solodit.cyfrin.io/issues/m-03-flash-loan-fee-is-incorrect-in-private-pool-contract-code4rena-caviar-caviar-private-pools-git}{033} }} & \SetCell[r=5]{l}{\textbf{\href{https://github.com/code-423n4/2023-04-caviar}{2023-04-caviar}}} & \SetCell[r=5]{l}{\textbf{5}} & Correct
& \xmark~ test\_FlashLoanFeeIncorrectlyScaled & \texttt{assertEq(expectedFlashFee/actualFlashFee,10**14)}\\
& & & Correct & \xmark~  test\_ChangeFeeVsFlashFeeDiscrepancy &
\texttt{assertEq(flashFeeAmount, 25)} \\
& & & Correct & \xmark~  test\_EconomicImpactOfVulnerability & \texttt{assertEq(actuallyPays, 2500)}\\
& & & Correct & \xmark~  test\_ExploitFlashLoanWithMinimalFee & \texttt{assertEq(feeReceived, 25)}\\
& & & Correct & \xmark~ test\_FlashLoanFeeIncorrectlyScaledWithERC20 &
\texttt{assertEq(actualFee, 25)}\\

\SetCell[r=2]{l}{\textbf{\href{https://solodit.cyfrin.io/issues/h-1-malicious-user-can-overtake-a-prefunded-auction-and-steal-the-deposited-funds-sherlock-axis-finance-git}{041}}} & \SetCell[r=2]{l}{\textbf{\href{https://github.com/sherlock-audit/2024-03-axis-finance}{2024-03-axis-finance} }} & \SetCell[r=2]{l}{\textbf{2}} & N/A
& \cmark~ test\_RootCause\_LotIdInitializedToZero & N/A\\
& & & Correct & \xmark~ test\_StealPrefundedTokensByOverwritingLot... &
%RoutingZero  
\texttt{assertEq(seller0After, ATTACKER)}\\

\SetCell[r=3]{l}{\textbf{\href{https://solodit.cyfrin.io/issues/m-03-zero-token-transfer-can-cause-a-potential-dos-in-cvxstaker-code4rena-xeth-xeth-versus-contest-git}{046}}} & \SetCell[r=3]{l}{\textbf{\href{https://github.com/code-423n4/2023-05-xeth}{2023-05-xeth} }} & \SetCell[r=3]{l}{\textbf{3}} & Correct
& \xmark~ test\_getReward\_RevertsOnZeroBalance & 
\texttt{vm.expectRevert("RevertOnZeroToken: zero amount")}\\
& & & Correct & \xmark~  test\_getReward\_BlocksLegitimateRewards & 
\texttt{vm.expectRevert("RevertOnZeroToken: zero amount")}\\
& & & Correct & \xmark~ test\_getReward\_PermanentDoS &
\texttt{vm.expectRevert("RevertOnZeroToken: zero amount")}\\

\SetCell[r=2]{l}{\textbf{\href{https://solodit.cyfrin.io/issues/m-01-contract-phinft1155-cant-be-paused-code4rena-phi-phi-git}{070}}} & \SetCell[r=2]{l}{\textbf{\href{https://github.com/code-423n4/2024-08-ph}{2024-08-ph} }} & \SetCell[r=2]{l}{\textbf{2}} & Correct
& \xmark~ test\_PauseDoesNotPreventTransfers &
\texttt{nft.safeTransferFrom(attacker, victim, tokenId, 1, "")};\\
& & & Correct & \xmark~  test\_PauseDoesNotPreventBatchTransfers &
\texttt{nft.safeBatchTransferFrom(.., .., tokenIds, amounts, "")}\\

\SetCell[r=3]{l}{\textbf{\href{https://solodit.cyfrin.io/issues/h-01-pumps-are-not-updated-in-the-shIAt-and-sync-functions-allowing-oracle-manipulation-code4rena-basin-basin-git}{091}}} & \SetCell[r=3]{l}{\textbf{\href{https://github.com/code-423n4/2023-07-basin}{2023-07-basin}}} & \SetCell[r=3]{l}{\textbf{3}} & N/A
& \cmark~ test\_correct\_behavior\_swap\_updates\_pump\_first & N/A\\
& & & Correct & \xmark~ test\_exploit\_sync\_oracle\_manipulation & \texttt{assertEq(pump.updateCount(), initialUpdateCount + 1}\\
& & & Correct & \xmark~ test\_exploit\_shift\_oracle\_manipulation &
\texttt{assertEq(pump.updateCount(), initialUpdateCount + 1}\\

\end{tblr}
\caption{Manual correctness analysis for cases with existing auditor PoCs. The table reports, for PoCs generated by \tool with \claude, the number of tests per PoC, their outcomes on the patched version (\xmark~ fail / \cmark~pass), and the exploit assertions associated with each failing test. A failing test on the patched version (\xmark) indicates that the patch mitigates the exploit.}
\label{tab:manual_rq2}
\end{table*}

% unsuccessful 
%008, fails RQ1
%015, Inconclusive already discussed
%018, Incorrect by Claude
%042, Fails RQ1
%048, Fails RQ1
%054, Incorrect
%077 Fails RQ1 MT

\paragraph{\textbf{Universal failure cases.}}\phantomsection\label{par:universal-failure-cases} Furthermore, we manually reviewed the cases in which \tool fails to generate a PoC across all three selected models. These cases are \href{https://solodit.cyfrin.io/issues/m-2-utilization-rate-multiplier-will-not-shIAt-IA-oracle-is-consulted-frequently-sherlock-cap-git}{\#042} and \href{https://solodit.cyfrin.io/issues/h-01-no-revert-on-transfer-erc20-tokens-can-be-drained-code4rena-cally-cally-contest-git}{\#054}.

The vulnerability in case \href{https://github.com/sherlock-audit/2025-07-cap-judging/issues/148}{\#042} is related to a rounding error in a rate multiplier.
\claude fails to generate a well-formed PoC because it reaches the maximum cost limit during generation. \glm fails in terms of logical correctness, as the generated PoC focuses on demonstrating the concrete mathematical miscalculation, decoupled from the actual contract logic; therefore, the PoC is not invalidated by the patch. 
Finally, \openai fails to correctly encode the security property: the PoC asserts that the rate changes across two different setups, while the vulnerability requires proving that the rate does \textit{not} change. Consequently, the generated exploit verifies the incorrect condition and does not expose the vulnerability. To summarize, all models fail to capture the code semantics of the problem.

Case \href{https://solodit.cyfrin.io/issues/h-01-no-revert-on-transfer-erc20-tokens-can-be-drained-code4rena-cally-cally-contest-git}{\#054} describes a subtle corner case in which a vault is configured to manage assets as if they were ERC-721 tokens, but is actually provided with an ERC-20 token. This mismatch in token type creates a deceptive situation: the contract follows transfer logic intended for a different asset standard and does not properly verify whether the ERC-20 transfer has succeeded, meaning that a failed transfer may go unnoticed and appear successful to users interacting with the vault. 
The exploit relies on this inconsistency between the declared and actual token type. 
However, instead of reproducing this specific scenario, the evaluated models focused on a different ERC-20–related issue, which is not referred to in the prompt.
We attribute this failure across all models to a vague vulnerability annotation that omits the critical ERC-721/ERC-20 mismatch.

Both failure cases illustrate that our evaluation criteria are strong for assessing the automatically generated PoCs, which verifies whether it exercises the vulnerability-related code path and whether it correctly captures the underlying vulnerability.

\begin{answerbox}{Answer to RQ2} \textbf{\rqtwo}
\tool's agentic architecture succeeds in generating logically correct exploits, producing nearly three times as many correct exploits as the best workflow baseline. 
The autonomous planning and tool-use capabilities enable \tool to express the described vulnerabilities in code, with assertions demonstrating the security flaw.
\tool clearly demonstrates superior exploit generation capabilities over non-agentic approaches.
\end{answerbox}

\subsection{RQ3 Results}\label{subsec:rq3-results}
Table~\ref{tab:RQ3_result} reports \tool:\claude's logical correctness across the three annotation levels on the nine samples where all three versions could be constructed (Section~\ref{subsec:rq3-methodology}). The levels span a wide range of detail: high-level annotations contain a median of approximately 30 words, detailed annotations 196, and procedural annotations 462.
The table uses the same error categories as RQ2; ill-formed cases from RQ1 are indicated by ``\textemdash''.

%\SetTblrInner{rowsep=1pt, colsep=2pt}

\begin{table}
\centering
\begin{tblr}{
    colspec = {cl ccc },
    hline{1,2,11} ={solid},
    vline{2,3,4,5}={solid},
    rows={font=\scriptsize},
    row{Z}={font=\bfseries},
}
\textbf{ID} & 
\textbf{Project} & 
\textbf{High-level} & 
\textbf{Detailed} & 
\textbf{Procedural} 
\\
    \href{https://solodit.cyfrin.io/issues/m-01-multicall-does-not-work-as-intended-code4rena-size-size-git}{001} &
    \href{https://github.com/code-423n4/2024-06-size}{2024-06-size} & 
     % Abstract
      \textemdash    &
     % Descriptive
     \success &  % test fails 
     % P
     \success   
     % assertion
    \\
    \href{https://solodit.cyfrin.io/issues/m-08-loss-of-funds-for-traders-due-to-accounting-error-in-royalty-calculations-code4rena-caviar-caviar-private-pools-git}{009} &
    \href{https://github.com/code-423n4/2023-04-caviar}{2023-04-caviar} &
    % A
    IC    &  % "executability": "compiles_and_passes", "logical": "test_passes"
    
    \success  & % test fails       
    
    \success    
    \\
    \href{https://solodit.cyfrin.io/issues/m-3-share-price-inflation-by-first-lp-er-enabling-dos-attacks-on-subsequent-buyshares-with-up-to-1001x-the-attacking-cost-sherlock-dodo-gsp-git}{020} & 
    \href{https://github.com/sherlock-audit/2023-12-dodo-gsp}{2023-12-dodo-gsp} &
    % A
    \textemdash    & % "executability": "compiles_but_fails",
    
    \textemdash    & % "executability": "compiles_but_fails",
    %P 
     \success      
    \\
    \href{https://solodit.cyfrin.io/issues/m-06-denial-of-service-contract-owner-could-block-users-from-withdrawing-their-strike-code4rena-putty-putty-contest-git}{032} &
    \href{https://github.com/code-423n4/2022-06-putty}{2022-06-putty} &
     \textemdash   & %    "executability": "does_not_compile",
    
     \textemdash   & %    "executability": "does_not_compile", 
    
     \textemdash     
    \\
    \href{https://solodit.cyfrin.io/issues/m-2-utilization-rate-multiplier-will-not-shift-if-oracle-is-consulted-frequently-sherlock-cap-git}{042} &
    \href{https://github.com/sherlock-audit/2025-07-cap}{2025-07-cap} &
    % A
      \textemdash   & % does not compile   
     \textemdash   & %    "executability": "does_not_compile",
    
     \textemdash     
    \\
    %%%
    \href{https://solodit.cyfrin.io/issues/h-01-royalty-receiver-can-drain-a-private-pool-code4rena-caviar-caviar-private-pools-git}{048} &
    \href{https://github.com/code-423n4/2023-04-caviar}{2023-04-caviar} &
    % A
      \textemdash   & %  "executability": "does_not_compile",
    
     \textemdash   & %    "executability": "does_not_compile", 
    
     \textemdash     
    \\
    \href{https://solodit.cyfrin.io/issues/h-08-player-can-mint-more-fighter-nfts-during-claim-of-rewards-by-leveraging-reentrancy-on-the-claimrewards-function-code4rena-ai-arena-ai-arena-git}{077} &
    \href{https://github.com/code-423n4/2024-02-ai-arena}{2024-02-ai-arena} &
    % A
    \success    & % "logical": "test_fails"
    
    \success    & % "logical": "test_fails"
    % P
    \textemdash    % NOTE verified that these results are accurate 
    \\
    \href{https://solodit.cyfrin.io/issues/h-01-pumps-are-not-updated-in-the-shift-and-sync-functions-allowing-oracle-manipulation-code4rena-basin-basin-git}{091} &
    \href{https://github.com/code-423n4/2023-07-basin}{2023-07-basin} &
    % A
     \textemdash   & %  "executability": "does_not_compile",
    
     \textemdash   &  % "executability": "compiles_but_fails", 
    %P
     \success     
    \\
    %%%%%%%%%%%%
    \href{https://solodit.cyfrin.io/issues/h-03-wp-h0-fake-balances-can-be-created-for-not-yet-existing-erc20-tokens-which-allows-attackers-to-set-traps-to-steal-funds-from-future-users-code4rena-cally-cally-contest-git}{098} &
    \href{https://github.com/code-423n4/2022-05-cally}{2022-05-cally} &
    % A
      \textemdash   & %  "executability": "does_not_compile",
      \textemdash   &  % "executability": "compiles_but_fails",    
    
     \success    
    \\
    \SetCell[c=2]{r}\textbf{Total Ill-formed (\textemdash)}&& 7 & 6 & 4\\
    \SetCell[c=2]{r}\textbf{Total Incorrect (IC)}    & & 1 & 0 & 0\\
    \SetCell[c=2]{r}\textbf{Total Inconclusive (IN)} & & 0 & 0 & 0\\
    \SetCell[c=2]{r}\textbf{Total Correct (\success)} & & 1 & 3 & 5\\
\end{tblr}
\centering
\caption{RQ3 Overview: Logical correctness dependence of annotation quality. RQ2 rerun with varying levels of detail in annotations using \claude. \textemdash{} indicates the case did not produce a well-formed PoC.} %\todo{Include the word, add a column to the right with assertion example}
\label{tab:RQ3_result}
\end{table}

For the least detailed annotation versions (high-level), \tool:Claude generated one correct PoC (\#007), and one incorrect case (\#009). %2024-02-ai-arena. 
When provided with detailed annotations, the number of correct PoCs increased to three (\#001, \#009, and \#077). 
%\#009 fails to produce a valid PoC for the more abstract high-level version. 
With the most detailed, procedural annotations, \tool:Claude produced five logically correct PoCs in total.

Overall, PoC quality increased with annotation detail in 5 of 9, demonstrating that richer vulnerability descriptions improve automated PoC exploit generation. 
%This result is actionable for auditors; they have an incentive to describe the problems well so that they can increase the likelihood of obtaining a PoC.
Sample \#077 represents the sole exception, where adding procedural information instead yielded negative effects. We attribute this to agentic randomness and hypothesize that a retry should yield a correct PoC.

To further illustrate these cases, we discuss representative case studies below.

\subsubsection{Overconstraining from Procedural Detail (\#077)}\label{sec:077}
%\subsubsection{Case Study of 2024-02-ai-arena (\#077)}
% We detailed this vulnerability in RQ1 (Section~\ref{sec:rq1-results}) where \tool:\claude struggled to produce a valid PoC. In this RQ, we further analyze \tool:\claude under different annotation qualities. 
AI Arena gamifies neural network training by allowing users to earn ERC-721 rewards. The vulnerability is a reentrancy in the \texttt{claimRewards} function, allowing a winning user to mint more NFTs than awarded.

All three annotation levels built comparable harnesses, yet only the abstract and descriptive runs finished with a logically correct PoC; the procedural input did not yield well-formed assertions.

The procedural prompt provides step-by-step instructions for solving the problem. \tool does reproduce the narrated exploit verbatim: it wires the attack to mint exactly six NFTs and enforces that outcome with \texttt{assertEq(nftsMinted, 6)}. Because \texttt{claimRewards} increments its internal \texttt{claimIndex} on every mint, including those triggered during reentrancy, the contract actually consumes more calldata slots than the scripted plan budgeted, and an array pre-sized for “six mints” were exhausted during execution, yielding a repeated array-out-of-bounds panic. The agent’s own trajectory (Figure~\ref{fig:077-proc-reason}) shows it rereading “mint six NFTs instead of three,” amplifying the story to ten mints, and chasing those counts instead of loosening the invariant.

% \begin{figure}[t]
%   \centering
%   \captionsetup{type=lstlisting}
%   \begin{tcolorbox}[
%       colback=gray!5,
%       colframe=black!75,
%       width=\textwidth,
%       arc=2mm,
%       boxrule=0.5pt,
%       left=6pt,
%       right=6pt,
%       top=6pt,
%       bottom=6pt]
%   {\small
%   \noindent\textbf{Trajectory \tool:\claude  (\#077 Procedural)}\\[2pt]
%   \emph{Assistant (seq.~40):} ``The annotation says roundId~3 should mint six NFTs instead of three... let me trace each step of that flow.''\\[4pt]
%   \emph{Edit (tool call):} Updates the comment block to read ``Total: 10 NFTs instead of 6,'' extending the story rather than relaxing the requirement.\\[4pt]
%   \emph{Assistant (seq.~82):} ``Array out of bounds! Because claimIndex continues across calls, the outer run needs indices 0,1,2 and the reentrant run needs 0,1... we're still passing the same array to both.''
%   }
%   \end{tcolorbox}
%   \caption{Key moments from the \#077 procedural run: the procedural description annotation keeps \tool chasing the narrated NFT counts instead of producing a minimal PoC.}
%   \label{fig:077-proc-reason}
% \end{figure}
\begin{figure*}[t]
  \centering
  \small
  \captionsetup{type=lstlisting}
  \begin{tcolorbox}[
      colback=blue!4,          % very light blue background
      colframe=blue!50!black,  % muted blue frame
      width=\textwidth,
      arc=2mm,
      boxrule=0.6pt,           % slightly softer border
      left=6pt,
      right=6pt,
      top=6pt,
      bottom=6pt]

  \textbf{PoC Trajectory: Claude Sonnet 4.5 (\#077, Procedural)}\\[4pt]

  \begin{tabularx}{\textwidth}
    {@{}>{\raggedleft\arraybackslash\footnotesize}p{0.7cm}
       p{1.5cm}
       @{\hspace{0.6em}} X@{}}

  [seq.~40] & \textsc{Assistant}: &
``The annotation says roundId~3 should mint six NFTs instead of three... let me trace each step of that flow.'' \\[3pt]

  [seq.~42] & \textsc{Tool}: &
  Applies a patch increasing the annotated mint total from 6 to 10 NFTs. \\[3pt]

  [seq.~82] & \textsc{Assistant}: &
 ``Array out of bounds! Because claimIndex continues across calls, the outer run needs indices 0,1,2 and the reentrant run needs 0,1... we're still passing the same array to both.''

  \end{tabularx}

  \end{tcolorbox}
  \caption{Extracted moments from the \#077 procedural run: the procedural description annotation overfits \tool's trajectory to an inaccurate scenario.}
  \label{fig:077-proc-reason}
\end{figure*}

By contrast, the descriptive and abstract prompts let the agent focus on the invariant that “the attacker obtains more NFTs than entitled”, so it.
(i) over-provisioned calldata to tolerate extra mints and (ii) asserted \texttt{assertGt} rather than an exact count.

In summary, the procedural guidance overconstrained the agent too much and failed, whereas higher-level descriptions left sufficient flexibility for autonomously solving the problem.

\subsubsection{Annotation Detail Enabling Correct Setup (\#009)}
%\subsubsection{Case Study of 2023-04-caviar (\#009)} 
This vulnerability occurs when NFT royalty recipients are set to \texttt{address(0)}, fees are deducted during trades but never delivered, permanently locking funds.
Triggering this issue requires a specific misconfiguration that is not present in standard NFTs: the PoC must explicitly create an NFT with a zero-address recipient to demonstrate the flaw.

The three different annotations produce different outcomes. For High Level, the PoC exploit is incorrect.
For the high-level annotation, its PoC reuses the standard Milady NFT, see Listing~\ref{lst:rq3_009}, left box line 3, which has a fixed, non-zero recipient. This prevents triggering the core \texttt{address(0)} condition, making the exploit logically incorrect.
On the other hand, when having the detailed annotation, the PoC exploit correctly deploys a custom \texttt{MisconfiguredNFT} (see line 10 in Listing~\ref{lst:rq3_009} on the right side) and sets the recipient to `address(0)`. This directly triggers the vulnerability and proves funds are lost (L15).
When fed with the procedural annotation, the PoC is also correct. Yet, when manually analyzing its code, we confirm that it is of higher quality, with more sanity checks and more realistic fee calculations.

This case shows that without explicit detail about the \texttt{address(0)} recipient condition, the tool cannot generate a well-formed exploit for this specific vulnerability.

\begin{figure*}[t]
\centering
\begin{minipage}[t]{0.48\textwidth}
\begin{lstlisting}[language=Solidity,
  basicstyle=\scriptsize\ttfamily,]
contract ExploitTest is Fixture {
    PrivatePool public privatePool;
    address nft = address(milady);
    // Hardcoded in Milady.royaltyInfo()
    address royaltyRecipient = address(0xbeefbeef);
    ...
    test_BuyRoyaltyInconsistency(
    ...
    // VULNERABILITY PROOF:
    // The buyer pays MORE than the quoted amount
    ... 
    assertEq(
        actualRoyaltyPaid,
        expectedTotalRoyalty,
        "Royalty recipient received the calculated royalty"
        );
    ) 
}
\end{lstlisting}
\end{minipage}
\hfill
\begin{minipage}[t]{0.48\textwidth}
%\textbf{\tool:GLM (custom helper)}\\[4pt]
\begin{lstlisting}[language=Solidity,
  basicstyle=\scriptsize\ttfamily,]
contract MisconfiguredNFT is ERC721, ERC2981 {}
  
contract ExploitTest is Fixture {
    PrivatePool public privatePool;
    MisconfiguredNFT public misconfiguredNFT;
    ...
 test_RoyaltyFeeLossOnBuy(){
    ...
  // recipient is address(0)
  misconfiguredNFT.setRoyaltyInfo(royaltyFeeRate, address(0)); 
    .. }
  // 3. Demonstrate the financial loss
  // The buyer paid expectedTotalRoyalty but it was never transferred to anyone
  // This amount is now stuck in the pool
  assertGt(expectedTotalRoyalty, 0, 
  "Royalty fee should be non-zero");
 }
\end{lstlisting}
\end{minipage}
\caption{\tool: exploit on finding 009 using abstract(left) and detailed (right) annotation .}
\label{lst:rq3_009}
\end{figure*}

\begin{answerbox}{Answer to RQ3} \textbf{\rqthree}
We observe a clear hierarchy in success rates: procedural > detailed > high-level annotations. This demonstrates that comprehensive vulnerability descriptions enable substantially better automated PoC generation, with even moderate technical detail (code snippets, explanations) providing benefit over minimal summaries.

\end{answerbox}

\section{Discussion}
\label{sec:discussion}

\subsection{Lessons Learned}
Our evaluation demonstrates that an LLM agent can produce logically correct PoC exploits for real-world smart contract vulnerabilities without human intervention. Beyond this core feasibility result, our experiments reveal three specific lessons.

\textbf{Lesson 1: Agent autonomy matters more than model capability.} \glm, the least capable model evaluated, produces zero well-formed PoCs under both Zero-shot and Workflow configurations, yet generates 16 with \tool. The ability to explore the codebase, reuse existing test infrastructure, and iteratively debug against compiler and runtime feedback enables PoC generation. This suggests that for PoC generation, enabling autonomous exploration with clear feedback signals yields greater returns than scaling model capacity alone.

\textbf{Lesson 2: Over-specified annotations may overconstrain the agent.} Procedural step-by-step annotations do not reliably improve results over detailed descriptions. In case \#077, procedural guidance overconstrained the agent into reproducing exact exploit parameters, causing repeated failures that higher-level descriptions avoided (Section~\ref{sec:077}). Across the nine samples evaluated with all three annotation levels, moderate technical detail consistently helps, but step-by-step exploitation instructions are not always beneficial. 
The practical implication is that auditors should provide technical detail about \textit{what} is broken and \textit{why}, while avoiding rigid procedural prescriptions that may conflict with the agent's own construction strategy.
%The practical implication is that auditors should describe \textit{what} is broken, but not necessarily define \textit{how} to exploit it \todo{Sofia: I disagree with the statement. Maybe to avoid overly detail description as they might give contradictory instructions? }.

\textbf{Lesson 3: Patch-based validation provides automated ground-truth for PoC correctness.} Our methodology of executing PoCs against developer-provided mitigation patches classified 44 of 50 well-formed PoCs automatically; only 6 required manual inspection due to patch-induced compilation changes. 
Additional manual analysis, including semantic comparison with auditor-written PoCs and review of universal failure cases, confirmed the soundness of the automated verdicts (Section~\ref{sec:rq2-results}). 
This methodology is reusable for any future work on PoC generation, making it a transferable contribution beyond \tool.

\subsection{Scope and Limitations}
\tool assumes as input a previously identified vulnerability and assists in generating a proof-of-concept (PoC) exploit for that reported issue. 
It does not assess the correctness of audit findings, the filtering of false positives, or the proving of the absence of vulnerabilities.
\tool does not automatically guarantee that a generated PoC fully demonstrates the intended vulnerability. 
Future work could integrate automatic verification oracles to validate PoC correctness against the vulnerability specification.

\subsection{Discussion of Patch-based Validation}
Our evaluation uses the developer-provided patch as an oracle for PoC correctness (Section~\ref{subsec:rq2-methodology}). This is because of the semantic relationship between patches and exploits: a correct patch must block all exploitable paths, while an exploit only needs to exercise a single path~\cite{arkin2005software}. This methodology comes with two risks regarding construct validity.

\textbf{Risk 1: Incomplete Patches.} If a patch does not fully fix the vulnerability, a correct PoC may still succeed on the patched code and be misclassified as incorrect in our evaluation. We mitigate this by collecting patches validated and adopted by the original protocol maintainers. Our manual analysis (Section~\ref{par:universal-failure-cases}) did not reveal incomplete patches, but this is not a formal guarantee.

\textbf{Risk 2: Coincidental Invalidation.} A patch may block the PoC without the PoC ever concretely exercising the vulnerable root cause. For example, if triggering a vulnerability requires steps A and B, a PoC reaching only step A could still be blocked by a patch that removes A, producing a false positive. We mitigate this risk by constructing a high-quality dataset where code changes are localized to the vulnerable region and do not contain behavioral divergence. Our manual analysis (Section~\ref{par:correctness-ground-truth-pocs}) did not reveal instances of coincidental blocking, but this is not a formal guarantee.

\subsection{Threats to Validity}\label{sec:threadstovalidity}
%\subsubsection{Internal Validity}
\noindent\emph{Baseline Design Space:} While we compare against two meaningful baselines, the design space for prompt techniques is infinite. Our baselines represent the closest alternatives to \tool's design, but different implementations might yield different results.

\noindent\emph{Resource Constraints:} Our experiments are limited by maximum tool calls and overall cost to ensure a fair comparison and practical feasibility. These constraints may affect the quality of the result, particularly for Claude Code, where most failures were due to these limits (in particular, cost). Higher experimental boundaries might improve performance.

\noindent\emph{Annotation Subjectivity:} Categorizing natural-language vulnerability descriptions into distinct levels involves inherent subjectivity. It is possible our three levels are not completely accurately representing each level on a sample comparison basis.

\noindent\emph{Data Leakage:} 12 of 23 vulnerabilities have existing public exploits that could be in training data. However, the performance difference across annotation types indicates that the tool reasons about new information rather than recalling memorized solutions.

%\subsubsection{External Validity}
\noindent\emph{Blockchain Ecosystem Specificity}: Our results may not generalize to different blockchain ecosystems. \tool is configured explicitly for Solidity-based protocols using the Foundry testing framework. We believe, however, that our results hold for other smart contract stacks.

\noindent\emph{Reproducibility Considerations:} The use of two proprietary closed models limits exact reproducibility across different environments. To mitigate this, we included an open-weight model (\glm).

\section{Related Work}\label{sec:related_work}

\subsection{Smart Contract Exploit Generation}
%\input{tables/exploit_tools}

%Intro paragraph about what started the research area.
%symbolic execution, early work
Early work focused on identifying specific vulnerability patterns in contract bytecode. Teether~\cite{teether} pioneered this area by symbolically analyzing bytecode to find transaction sequences that could lead to malicious control flows, such as unauthorized transfers or code execution. Building on this, Maian~\cite{maian} automated the detection of three critical vulnerability types (leaking, locked, and suicidal contracts) by symbolically exploring flawed paths and concretely validating them on a forked blockchain.
A significant advancement was introduced by SmartScopy~\cite{feng2019precise} by identifying vulnerable transaction sequences and automatically constructing a concrete adversarial contract capable of triggering the vulnerability, effectively bridging the gap between vulnerability discovery and weaponized exploit generation. 
Exgen~\cite{Jin2023ExGen:Vulnerabilities} extended this line of work by introducing Partially-ordered Transactional Sets (PTS) to model complex, non-linear transaction dependencies, enabling the generation of sophisticated multi-transaction exploits for vulnerabilities like reentrancy and integer overflows.

%fuzzing
To improve scalability and avoid the path explosion problem of pure symbolic execution, subsequent research turned to fuzzing. EthPloit~\cite{Kontogiannis2020ETHPLOIT:Contracts} combined static analysis with fuzzing to discover exploitation patterns. ContraMaster~\cite{Wang2022Oracle-SupportedContracts} employed a two-phase architecture with a fuzzer to generate transactions and an instrumented EVM to validate behavioral violations. More advanced fuzzing techniques incorporate reinforcement learning. 
MADFuzz~\cite{mad_fuzz} uses Multi-agent Reinforcement Learning to guide the generation of effective transaction sequences. EF\/CF~\cite{efcf2023} prioritizes performance by translating EVM bytecode to C++, enabling high-speed fuzzing to discover and validate exploit sequences on a local chain.

% LLM and Exploit Synthesis
Recent work leverages the generative capabilities of LLMs for exploit synthesis. XploGen~\cite{xplogen_eshgie} utilizes an LLM guided by a formal DCR-graph oracle to inject business logic vulnerabilities and generate JavaScript exploit sequences. However, its reliance on hard-to-obtain formal specifications and its reported low exploit completion rate (29\%) limit its practical applicability. 
TracExp~\cite{su2026transactions} targets \textit{post-incident} analysis by synthesizing PoCs from observed on-chain attack traces, combining trace-driven decompilation with LLM-based code generation.
AdvScanner~\cite{advscanner} combines LLMs with static analysis in a feedback loop to generate adversarial contracts specifically for reentrancy vulnerabilities, but it does not produce end-to-end exploit tests.
VeriExploit~\cite{wei2025veriexploit} combines formal verification counterexamples with LLM-driven synthesis to create attacker contracts, but is limited to small single-contract programs due to the scalability constraints of symbolic execution.
Du and Tang~\cite{du2024evaluation} evaluate GPT-4’s auditing capabilities on 30 smart contracts with artificially injected vulnerabilities, reporting limited vulnerability detection performance but promising PoC writing ability when vulnerability hints are provided.
ReX~\cite{xiao2025prompt2pwn} uses LLMs to generate Solidity exploit code, which is then executed and tested within the Foundry framework. However, it follows a fixed pipeline without autonomous codebase exploration, which prevents reasoning across multi-contract codebases. Similarly, A1~\cite{gervais2025ai} presents an agentic framework where an LLM-powered agent equipped with various tools autonomously monitors and exploits on-chain contracts for profit. However, its scope is restricted to vulnerabilities with direct monetary extraction, excluding access control violations, denial-of-service attacks, and invariant breaches that are common in real-world audits.
SmartPoC~\cite{chen2025smartpoc} converts static-analysis findings into PoCs through context slicing, iterative repair, and LLM-assessed state-differential checks.\tool instead takes auditor-written descriptions as input, autonomously explores real-world multi-contract codebases, and validates correctness deterministically against ground-truth mitigation patches.

%Domain specific - more like flash loans,  and defi logic vuln
Another line of work targets vulnerabilities within particular application domains. 
FORAY~\cite{foray2024} focuses on discovering complex business logic vulnerabilities in DeFi protocols, such as price manipulation and flash loan attacks. It models protocol interactions as a Token Flow Graph and frames exploit generation as a graph search problem, synthesizing executable exploit contracts that chain multiple protocol functions.  
FlashSyn \cite{flashsyn2024} synthesizes flash-loan attacks by modeling DeFi protocols and using counterexample-guided refinement to generate exploit sequences.
CPMMX~\cite{Han2025AutomatedMakers} focuses on Constant Product Market Makers (CPMMs). It defines economic invariants for DEXes and uses fuzzing to discover transaction sequences that break these invariants profitably, often uncovering underlying vulnerabilities, such as logic errors. 
Osprey~\cite{approveonce_osprey_2025} addresses a critical vulnerability in the ERC-20 ecosystem, the Approved-Controllable-TransferFrom (ACT) flaw. It employs symbolic execution to detect vulnerabilities in deployed bytecode and generates exploit payloads to steal tokens from users who have approved a vulnerable contract.
Heimdallr~\cite{hu2026effective} proposes an agentic LLM-based framework for pre-deployment smart contract auditing that focuses on detecting business-logic vulnerabilities.

%\subsection{Audit Agents}

%\todo{Combining fine-tuning and llm-based agents for intuitive smart contract auditing with justifications}

%iAudit~\cite{iAudit}  is one of the first approaches at LLMs for automated smart contract auditing. It proposes a Prompting workflow  architecture in which 4 roles are defined: Detector, Reasoner, Ranker and Critic. In combinations this different steps identify a vulnerability based on generated and selected ``reasons'', and improved by ranking and criticizing the initial roles actions.
%While the concept of PoC is included in the prompts, it refers to what we describe as Procedural in this paper and no executable verification is performed.
%LLM-SmartAudit\cite{smartaudit} automates smart contract auditing using a multi-agent system where specialized LLMs collaborate to mimic a professional security team. It significantly outperforms both general-purpose LLMs and specialized tools like GPTScan and PropertyGPT
%\todo{Advanced Smart Contract Vulnerability Detection via LLM-Powered Multi-Agent Systems}

\subsection{Agentic AI for Offensive Security}\label{sec:related-offsec}
The success of LLMs in the code domain \cite{chen2021evaluating} has been leveraged to strengthen software security, particularly for automating penetration testing and vulnerability exploitation.

% 1. Early LLM-based systems are control loops
Early work established the paradigm of LLM-driven control loops for automating hacking. While PentestGPT \cite{deng2024pentestgpt} assists human testers through command suggestion (non-agentic), Happe and Cito \cite{happe2023pwnd} introduced autonomous planning and shell command execution. This modular control loop approach was refined in systems like PenHeal \cite{huang2023penheal} and AutoAttacker \cite{xu2024autoattacker}. Unlike these fixed-workflow systems, \tool{} autonomously plans and generates PoC exploits without predefined control loops.

% 2. realistic testbeds and autonomy
Recent efforts have progressed from developing control loops to testing autonomy in realistic settings. CTFs have emerged as a popular evaluation testbed for penetration testing due to their well-defined scope and success criteria. Shao et al. \cite{shao2024nyu} establish a key benchmark with 200 CTF challenges, catalyzing efforts to improve LLMs' hacking capabilities, such as Abramovich et al.'s \cite{abramovich2025enigma} EniGMA agent. Beyond CTFs, AutoPenBench \cite{gioacchini_autopenbench_2024} introduced a benchmark for penetration testing on real-world CVEs planted in virtual machines. Concurrently, research has considered multi-host network environments. Happe and Cito \cite{happe2025enterprise} propose Cochise and evaluate it on a 5-host Active Directory network, while Singer et al. \cite{singer2025feasibility} developed the Incalmo agent and a custom benchmark of 10 emulated networks. Sapia and Böhme~\cite{sapia2026scaling} target scaling automated security testing by using an LLM agent to autonomously handle the complex setup that real-world software requires before deep functionality becomes testable, then amplifying reached states with in-vivo fuzzing.

Similar to these efforts, which increasingly emphasize end-to-end automation in realistic settings, our work evaluates against real-world smart contract projects with confirmed, high-impact vulnerabilities, simulating the scenario faced by auditors when writing PoCs for their audits.

% 3. SAFETY
The proliferation of capable offensive agents has sparked a parallel line of research focused on evaluating their dual-use risks and real-world efficacy. Several dedicated benchmarks have been developed for this purpose, for example, Zhang et al. \cite{zhang2024cybench} and Dawson et al. \cite{dawson2025airtbench} create CTF-based benchmarks to scaffold LLM safety evaluations. 
Wei et al. \cite{wei2025dynamic} demonstrate that simple modifications to a baseline agent scaffold, such as repeated sampling, can significantly enhance offensive success rates. 
% SAFETY
Major LLM providers use such benchmarks for internal evaluations of their frontier models' offensive cybersecurity capabilities. The review of GPT-OSS \cite{agarwal2025gpt}, an OpenAI large open-source model, includes evaluations on CTF challenges and cyber range exercises. 

Our work is also dual-use and can be used for hacking real smart contracts in the field.
To mitigate risks, our evaluation is conducted under controlled conditions, excluding access to the mainnet, and our prototype is provided only upon request.

\section{Conclusion}
\label{sec:conclusion}
This paper introduced \tool, an agentic framework that automates the generation of PoC exploits for smart contract vulnerabilities. Our evaluation demonstrated that \tool's autonomous, tool-augmented approach significantly outperforms the baselines of Zero-shot and Workflow Prompting.
We demonstrated that \tool can successfully produce well-formed and logically correct exploits. 
By transforming natural-language vulnerability descriptions into executable tests, \tool addresses a critical bottleneck in the smart contract auditing process: the difficulty of obtaining PoCs. \tool provides auditors with verifiable evidence to strengthen their reports. It gives developers immediate test cases to understand and fix security flaws, thereby enhancing the overall security of smart contracts practically and cost-effectively.

% \section{Acknowledgement}
\section*{Acknowledgements}
This work was supported by the Centre for Cyber Defence and Information Security (CDIS), the WASP program funded by Knut and Alice Wallenberg Foundation, and by the Swedish Foundation for Strategic Research (SSF). Some computation was enabled by resources provided by the National Academic Infrastructure for Supercomputing in Sweden (NAISS).
We thank Linus Svensson for sharing his master's thesis work on automated benchmark generation. His scripts served as an initial foundation for the development of \dataset. 
We thank Dwellir for giving us access to their RPC infrastructure under academic license.

% \balance
\bibliographystyle{plain}
\bibliography{main}

\appendix
\section{Development Dataset}\label{app:dev-dataset}
The development dataset consists of 9 samples similar to those in \dataset, these are listed in Table~\ref{tab:dev-dataset}.

\begin{table}[h!]
\centering
\caption{Development Dataset}
\label{tab:dev-dataset}
\scriptsize
\begin{tabular}{@{}llll@{}}
\toprule
\textbf{Project} & \textbf{Vulnerability Type} & \textbf{Main Contract} & \textbf{Platform} \\ 
\midrule
2023-04-caviar & Flash Loan & PrivatePool.sol & Code4rena \\
2023-07-lens & Logical Error & FollowNFT.sol & Code4rena \\
2023-07-pooltogether & Access Control & Vault.sol & Code4rena \\
2024-02-ai-arena & Reentrancy & MergingPool.sol & Code4rena \\
2024-06-size & Logical Error & Multicall.sol & Code4rena \\
2024-06-vultisig & Flash Loan & ILOPool.sol & Code4rena \\
2025-01-iq-ai & Logical Error & TokenGovernor.sol & Code4rena \\
2025-01-liquid-ron & Access Control & LiquidRon.sol & Code4rena \\
2024-06-union-finance-update-2 & Access Control & VouchFaucet.sol & Sherlock \\
\bottomrule
\end{tabular}%

\end{table}

\section{\tool System Prompt}\label{app:system-prompt}
The full \tool system prompt is detailed in Figure~\ref{fig:system-prompt}.
\begin{figure*}[h!]
\centering
\begin{tcolorbox}[
    colback=gray!5,
    colframe=black!75,
    width=\textwidth,
    arc=2mm,
    boxrule=0.5pt,
    left=6pt,
    right=6pt,
    top=6pt,
    bottom=6pt
]
{\footnotesize
\noindent\textbf{System Prompt:}

\medskip
\noindent
You are an expert smart contract security testing specialist. Generate executable Proof-of-Concept (PoC) exploits demonstrating vulnerabilities using Foundry.

\medskip
\noindent\textbf{PoC Explainability.}
Write exploits as executable demonstrations that clearly prove the vulnerability. Include detailed comments documenting each attack step, the vulnerability being exploited, and why the exploit succeeds. The PoC must be self-explanatory to security auditors.

\medskip
\noindent\textbf{Vulnerability Analysis.}
Parse the vulnerability description (annotation) and analyze the vulnerability type, affected code sections, and potential impact. Analyze the contract logic to understand the root cause before developing exploits.

\medskip
\noindent\textbf{Testing Framework Guidelines.}
Use Foundry exclusively for testing. Implement proper \texttt{setUp()} functions with realistic contract states: i.e. initializing contracts with typical production values (reasonable token balances, realistic timestamps, standard protocol roles assigned). Utilize Foundry cheatcodes for test control: \texttt{vm.prank()} for identity switching, \texttt{vm.deal()} for ETH funding, \texttt{vm.warp()} for time manipulation, \texttt{vm.expectRevert()} for failure testing. Structure tests following Foundry conventions with clear test function names prefixed with \texttt{test}.

\medskip
\noindent\textbf{PoC Executability.}
Ensure all generated code compiles successfully with the specified Solidity version. Verify that tests pass (exploits vulnerability) when the vulnerability exists and fail when properly patched. Use \texttt{forge compile} and \texttt{forge test} to validate. Resolve all compilation errors, import issues, and version conflicts while preserving original contract logic.

\medskip
\noindent\textbf{Iterative Refinement.}
Debug compilation errors, test failures, and logical inconsistencies systematically using forge output and detailed error messages. For import path errors, check 1-2 existing test files to identify the correct pattern. Continuously improve until tests compile, execute successfully, and accurately demonstrate the vulnerability. If stuck on the same technical issue for >3 attempts, shift to a minimal working demonstration—proving the vulnerability exists matters more than perfect test coverage or setup complexity.

\medskip
\noindent\textbf{Exploit Soundness.}
Ensure exploits logically reflect the described vulnerability. The attack vector must accurately represent the security issue. Avoid false positives—exploits should fail if the vulnerability is fixed. Verify that the PoC demonstrates the actual impact described in the vulnerability description (annotation).

\medskip
\noindent\textbf{Exploit Quality.}
Keep PoCs minimal and focused. Write only the test file—never modify contracts under test or the original codebase. Reuse existing test infrastructure when available. Create helper contracts or mocks only when the exploit requires them. Avoid assumptions about undocumented contract behavior.}
\end{tcolorbox}
\caption{System prompt for PoC generation agent.}
\label{fig:system-prompt}
\end{figure*}

\section{Annotation Level Distribution}\label{app:annotation-levels}

Table~\ref{tab:annotation-levels} summarizes which annotation levels are available for each project. Of the 23 samples, 9 contain all three levels (high-level, detailed, and procedural), 12 contain only the first two, and 2 (\#018, \#049) contain only high-level and procedural annotations. The nine samples with complete coverage form the basis of the RQ3 analysis in Section~\ref{subsec:rq3-results}.

\begin{table}[h]
\centering
\scriptsize
\caption{Annotation Levels Extracted from original annotation.}
\label{tab:annotation-levels}
\begin{tabular}{clcccc}
\toprule

\textbf{ID} & \textbf{Project} & \textbf{Abstract} & \textbf{Descriptive} & \textbf{Procedural} & \textbf{All} \\
\midrule
001 & 2024-06-size & \checkmark & \checkmark & \checkmark & \checkmark \\
003 & 2023-07-pooltogether & \checkmark & \checkmark & & \\
008 & 2023-09-centrifuge & \checkmark & \checkmark & & \\
009 & 2023-04-caviar & \checkmark & \checkmark & \checkmark & \checkmark \\
015 & 2023-07-pooltogether & \checkmark & \checkmark & & \\
018 & 2023-04-caviar & \checkmark & & \checkmark & \\
020 & 2023-12-dodo-gsp & \checkmark & \checkmark & \checkmark & \checkmark \\
032 & 2022-06-putty & \checkmark & \checkmark & \checkmark & \checkmark \\
033 & 2023-04-caviar & \checkmark & \checkmark & & \\
039 & 2024-03-axis-finance & \checkmark & \checkmark & & \\
041 & 2024-03-axis-finance & \checkmark & \checkmark & & \\
042 & 2025-07-cap & \checkmark & \checkmark & \checkmark & \checkmark \\
046 & 2023-05-xeth & \checkmark & \checkmark & & \\
048 & 2023-04-caviar & \checkmark & \checkmark & \checkmark & \checkmark \\
049 & 2023-08-cooler & \checkmark & & \checkmark & \\
051 & 2023-09-centrifuge & \checkmark & \checkmark & & \\
054 & 2022-05-cally & \checkmark & \checkmark & & \\
058 & 2022-06-putty & \checkmark & \checkmark & & \\
066 & 2023-11-kelp & \checkmark & \checkmark & & \\
070 & 2024-08-ph & \checkmark & \checkmark & & \\
077 & 2024-02-ai-arena & \checkmark & \checkmark & \checkmark & \checkmark \\
091 & 2023-07-basin & \checkmark & \checkmark & \checkmark & \checkmark \\
098 & 2022-05-cally & \checkmark & \checkmark & \checkmark & \checkmark \\
\midrule
\textbf{Total} & & \textbf{23} & \textbf{21} & \textbf{11} & \textbf{9} \\
\bottomrule
\end{tabular}
\end{table}

\end{document}